\documentclass[aps,pre, floatfix,superscriptaddress]{revtex4}

{\begin{list}{}%
         {\setlength{\leftmargin}{#1}}%
         \item[]%
}
{\end{list}}

\usepackage{amssymb, amsmath, amsfonts}
\usepackage{bm}
\usepackage{multirow}
\usepackage{graphicx}
\include{subfigure}

\usepackage[normalem]{ulem} 
\usepackage[usenames,dvipsnames]{color}

\usepackage{subfigure}
\graphicspath{ {images/}{./Tex//} }
\begin{document}

\title{Predicting the dynamics of bacterial growth inhibition by ribosome-targeting antibiotics}

\date{December 2016\qquad *: Equal contribution}

\providecommand{\red}[1]{\textcolor{red}{#1}}

\providecommand{\green}[1]{\textcolor{green}{#1}}

\author{Philip Greulich$^*$}
\address{Mathematical Sciences, University of Southampton, Highfield Campus, SO17 1BJ, United Kingdom}
\address{Institute for Life Sciences, University of Southampton, Highfield Campus, SO17 1BJ, United Kingdom}
\author{Jakub Dole\v{z}al$^*$}
\address{School of Physics and Astronomy, University of Edinburgh, Peter Guthrie Tait Road, Edinburgh EH9 3FD, United Kingdom}
\author {Matthew Scott}
\address{Department of Applied Mathematics, University of Waterloo, Waterloo, ON, Canada}
\author{Martin R. Evans}
\address{School of Physics and Astronomy, University of Edinburgh, Peter Guthrie Tait Road, Edinburgh EH9 3FD, United Kingdom}
\author{Rosalind J. Allen}
\address{School of Physics and Astronomy, University of Edinburgh, Peter Guthrie Tait Road, Edinburgh EH9 3FD, United Kingdom}
\address{Centre for Synthetic and Systems Biology, The University of Edinburgh, Edinburgh, United Kingdom}

\noindent

\begin{abstract}
Understanding how antibiotics inhibit bacteria can help to reduce antibiotic use and hence avoid antimicrobial resistance -- yet few theoretical models exist for bacterial growth inhibition by a  clinically relevant antibiotic treatment regimen. In particular, in the clinic, antibiotic treatment is time dependent.  Here, we use a recently-developed model to obtain predictions for the dynamical response of a bacterial cell to a time-dependent dose of ribosome-targeting antibiotic. Our results  depend strongly on whether  the antibiotic shows reversible transport and/or low-affinity ribosome binding (``low-affinity antibiotic'') or, in contrast, irreversible transport and/or high affinity ribosome binding (``high-affinity antibiotic''). For low-affinity antibiotics, our model predicts that growth inhibition depends on the duration of the antibiotic pulse, with a transient period of very fast growth  following removal of the antibiotic. For high-affinity antibiotics, growth inhibition depends on peak dosage rather than dose duration, and the model predicts a pronounced post-antibiotic effect, due to hysteresis, in which growth can be suppressed for long times after the antibiotic dose has ended. These predictions are experimentally testable and may be of clinical significance.

\end{abstract}

\maketitle

\section{Introduction}

Modern clinical practice relies on the use of antibiotics to combat bacterial infections, yet our knowledge of how antibiotics inhibit bacteria is surprisingly incomplete.  In particular, even though the molecular processes  that are targeted by antibiotics are generally known \cite{franklin}, mathematical models for bacterial growth rate as a function of antibiotic concentration are generally lacking. Development of such models can help to combat antimicrobial resistance, because they are a crucial input for the optimisation of dosing regimes \cite{baym,Kishoney_Synergy}, and also for models of how bacteria evolve resistance to antibiotics \cite{allen_2016,antibiotic_gradients,gradients_hermsen,deris}. In recent work, models have been proposed that can predict the steady-state response of bacterial growth to a fixed antibiotic concentration \cite{greulich_2015,deris,Kishoney_Synergy,chevereau_2015}. In the clinic, however, the antibiotic concentration to which an infection is exposed is time-varying. In this paper, we present theoretical predictions for the dynamical changes in bacterial growth rate in response to a time-varying concentration of a ribosome-targeting antibiotic. Our analysis predicts qualitative, and potentially clinically relevant, differences in the dynamical response of bacterial growth to antibiotic treatment, depending on the molecular parameters for antibiotic-ribosome binding and transport of antibiotic across the bacterial cell boundary. \\

Measurements of bacterial growth inhibition by antibiotics often focus on exponentially growing bacterial populations, whose population dynamics can be described by $N(t) \sim \exp{(\lambda t)}$, where $N(t)$ is the number of bacteria at time $t$ and $\lambda$ is the growth rate \cite{growth_bacteria_book}. For bacteria that are grown in the presence of a sub-lethal concentration of antibiotic, the growth rate $\lambda$ is expected to decrease as the antibiotic concentration $a_{\rm ex}$ increases. The form of the growth-inhibition function $\lambda(a_{\rm ex})$ may depend on the antibiotic \cite{greenwood,greulich_2015}, the bacterial strain \cite{greenwood,deris}, the growth medium \cite{greulich_2015} and the presence of other antibiotics \cite{bollenbach_2015}. The concentration of antibiotic required to halve the growth rate, known as the IC$_{50}$, provides one way to quantify the efficacy of  an antibiotic.\\

Many antibiotics target bacterial ribosomes. Ribosomes are multi-component, molecular machines which carry out protein synthesis -- a function that is crucial for growth. Different ribosome-targeting antibiotics can bind to different components of the bacterial ribosome and inhibit different steps in protein synthesis \cite{wilson_2014}.  In recent experimental and theoretical work \cite{greulich_2015}, Greulich {\em et al} suggested that ribosome-targeting antibiotics can be classified according to key features of their growth-inhibition curves  $\lambda(a_{\rm ex})$ and the dependence of their IC$_{50}$ values on the bacterial growth medium. 
Remarkably, this work showed that contrasting experimental observations for the growth-medium dependent efficacy of different ribosome-targeting antibiotics can be reproduced by a simple mathematical model that takes account of the molecular processes of antibiotic-ribosome binding and antibiotic transport across the cell boundary, as well as the physiological processes of cell growth and ribosome synthesis \cite{greulich_2015}. In the model, the latter processes are  coupled to the state of the cell's ribosome pool via empirical ``growth laws'' \cite{scott_science,Scott2011}: experimentally-established mathematical relations that describe how a bacterial cell balances the production of new ribosomes and of other proteins, depending on its growth rate. Ref. \cite{greulich_2015} focuses on the response to a fixed (time-invariant) antibiotic concentration; here we subject the same model to time-dependent antibiotic perturbations.\\

In a clinical situation, during antibiotic treatment a bacterial infection experiences a time-varying local concentration of antibiotic \cite{greenwood}.  Pharmacokinetic curves describe the time-dependent local antibiotic concentration at an infection site; these curves show a peak as the antibiotic concentration initially increases following ingestion and transport to the active site and later decreases due to possible metabolism within the body and excretion \cite{greenwood,tozer}. Pharmacodynamics attempts to link these curves to the efficacy of antibiotic action \cite{greenwood,tozer}. In particular, some dosing protocols are designed to maximise the peak concentration, whereas others aim to maximise the time at which the concentration is maintained above a certain threshold, or, alternatively, the area of the curve which is above a threshold \cite{greenwood}. Importantly, for some antibiotics, activity can persist for some time after the antibiotic is removed: this is known as the post-antibiotic effect \cite{mackenzie_1993,bundtzen_1981} and occurs for a variety of antibiotics including ribosome-targeting aminoglycosides \cite{isaksson_1988,bundtzen_1981}. The mechanisms behind this effect are unknown but may include slow recovery after reversible damage to cell structures, slow removal of the antibiotic from its binding site, and the need to synthesize new enzymes before a bacterium can resume growth \cite{mackenzie_1993}.\\

In this paper, we use the mathematical model of ribosome-targeting antibiotics introduced in Ref. \cite{greulich_2015} to make dynamical predictions for the response of bacterial growth rate to a time-varying antibiotic concentration. The model predicts qualitatively different dynamical responses for ribosome-targeting antibiotics that bind to ribosomes with high affinity and/or are transported into the cell irreversibly (``high-affinity antibiotics''), as opposed to antibiotics that bind with low affinity and/or are transported reversibly (``low-affinity antibiotics'') . Our results reproduce some known pharmodynamic phenomena, such as a post-antibiotic effect for high-affinity ribosome-targeting antibiotics. Our model also predicts new phenomena, including a transient increase in growth rate upon removal of a low-affinity ribosome-targeting antibiotic. We suggest ways in which these predictions could be tested experimentally and comment on their potential clinical relevance.

\section*{Background: Mathematical model for the action of ribosome-targeting antibiotics}

\begin{figure}
\centering
\includegraphics[width=0.4\linewidth]{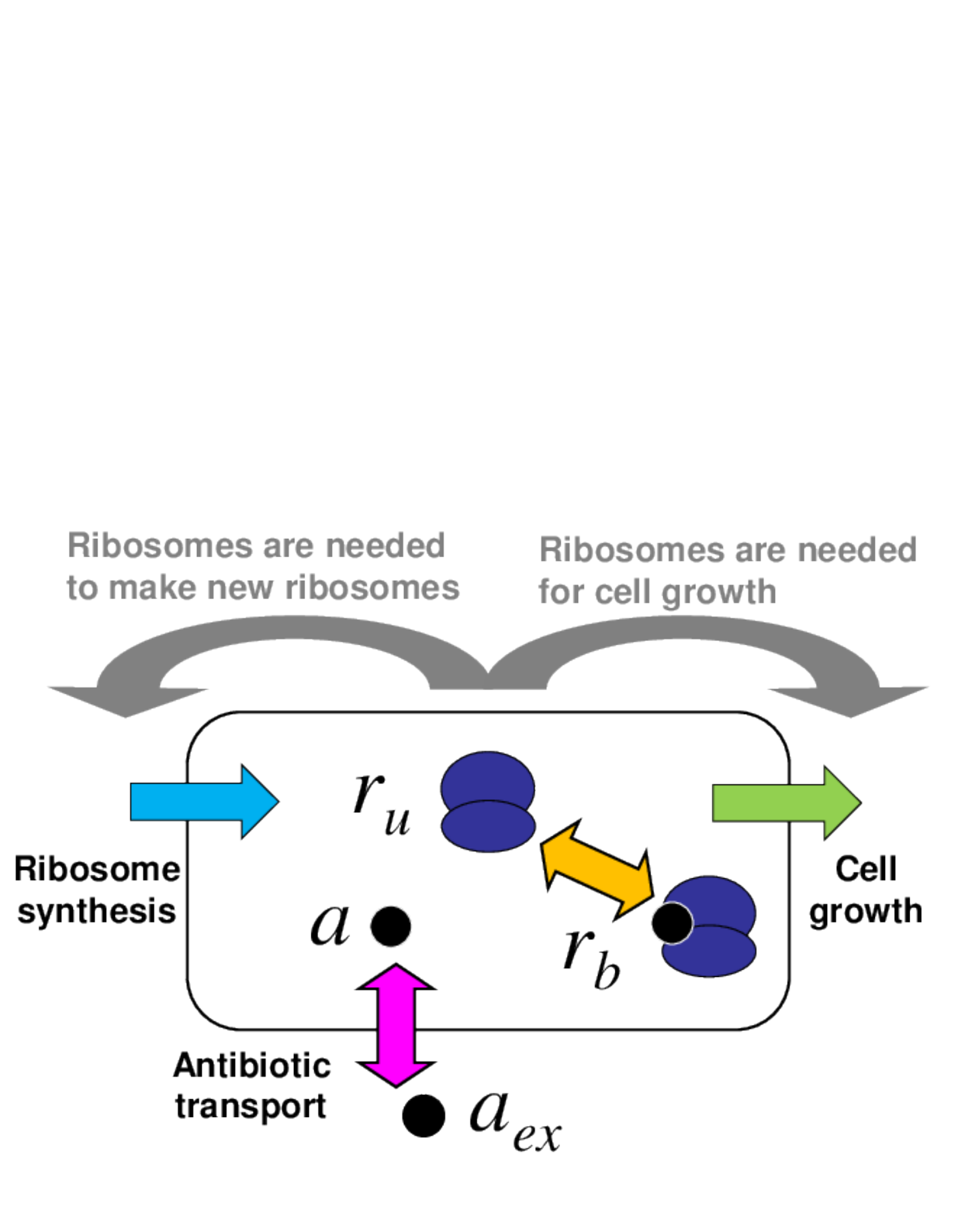}
\caption{Schematic illustration of the model. The bacterial cell is modelled as a well-mixed vessel, containing ribosomes (dark blue) which may be free or bound by antibiotic. Antibiotic molecules (black circles) can be transported into or out of the cell (pink arrow) and can bind to or dissociate from ribosomes (orange arrow). The model also includes cell growth (green) and ribosome synthesis (light blue), both of which are coupled to the state of the cell (these couplings are illustrated by the grey arrows).}
\label{fig:model}
\end{figure}

We first provide a brief description of the model which was introduced in Ref. \cite{greulich_2015}. As illustrated in Figure \ref{fig:model}, the model describes a bacterial cell as homogeneous mixture of ribosomes and antibiotic molecules, which can bind reversibly to the ribosomes. The variables of the model are the concentrations of bound and unbound ribosomes, $r_{\rm b}$ and $r_{\rm u}$ respectively, and the concentration of intracellular antibiotic $a$. The model consists of the following set of equations for the dynamics of the concentrations:
\begin{eqnarray}
\dot{a} &=&  - F(a,r_{\rm u},r_{\rm b})  - 
\lambda a + P_{\rm in} a_{\rm ex}  - P_{\rm out} a,
\label{eq:antibDyn}\\
\dot{r_{\rm u}} &=&  -  F(a,r_{\rm u},r_{\rm b})  - \lambda r_{\rm u}  + 
s,
\label{eq:ruDyn}\\
\dot{r_{\rm b}} &=& F(a,r_{\rm u},r_{\rm b}) - \lambda r_{\rm b} .
\label{eq:rbDyn}
\end{eqnarray}
Here, $F(a,r_{\rm u},r_{\rm b}) \equiv k_{\rm on} a\left( {r_{\rm u}  - r_{\min } } \right) - k_{\rm off} r_{\rm b}$ describes the antibiotic-ribosome binding / unbinding kinetics. $P_{\rm in}$ and $P_{\rm out}$ are rate constants (or permeabilities) for antibiotic transport into and out of the cell (both assumed to be linear processes), and  $a_{\rm ex}(t)$ is the time-varying external antibiotic concentration, which can be controlled in a lab experiment, or is set by the dosing regime in a clinical scenario. The bacterial growth rate is denoted by $\lambda$ and the terms $-\lambda a$, $- \lambda r_{\rm u}$ and $- \lambda r_{\rm b}$ describe dilution of the cell contents by growth. Finally, $s$ is the rate of synthesis of new ribosomes;  $s$ is a function of the state of the system and will be discussed below.\\


To complete the model, we also need to describe  how the growth rate $\lambda$ and the ribosome synthesis rate $s$ depend on the state of the system (grey arrows in Fig. \ref{fig:model}). To do this, we use the empirically determined ``growth laws'' of Scott {\em et al.} \cite{scott_science}. The growth rate $\lambda$ is taken to be linearly related to the concentration of unbound ribosomes $r_{\rm u}$: 
\begin{equation}
r_{\rm{u}}  = r_{\min }  + \frac{\lambda }
{{\kappa _t }}.
\label{eq:1stGL}
\end{equation}
Relation (\ref{eq:1stGL}) is based on experimental measurements for bacteria growing exponentially on a range of nutrients, in the absence of antibiotic \cite{scott_science,bremer_dennis}; we assume here, as in Ref. \cite{greulich_2015}, that it also holds in the presence of antibiotic  \footnote{The conversion between the units of fraction of cell mass used in Ref. \cite{scott_science} and ribosome concentration is discussed in the Supplementary Material of Ref. \cite{greulich_2015}.}. The constant $\kappa _t=6.1\times 10^{-2}\mu$M\,h$^{-1}$ is the translation rate of the ribosomes and the constant $ r_{\min }=19.3 \mu$M is believed to arise from an inactive pool of ribosomes which may be waiting to initiate translation or stalled during translation \cite{scott_science,Klumpp2013}. We have implicitly assumed that these inactive ribosomes do not bind antibiotic, through our definition of the binding function  $F(a,r_{\rm u},r_{\rm b})$ \footnote{The assumption that inactive ribosomes do not bind antibiotic simplifies the mathematical analysis; numerical investigations confirm that the qualitative behaviour of the model is the same if the inactive ribosomes are allowed to bind to antibiotic.}. \\

 The ribosome synthesis rate $s$ can also be deduced from experimental measurements, for bacteria growing exponentially. Ribosome synthesis is regulated such that inhibition of ribosome action by antibiotic causes the cell to increase ribosome production \cite{ribsynth_rev2,ribsynth_rev3,scott_2014}. Measuring the total ribosome content as a function of (decreasing) growth rate upon addition of a ribosome-targeting antibiotic, Scott {\em et al.} observed a linear relation between total ribosome content $r_{\rm tot}$ and growth rate \cite{scott_science}: 
\begin{equation}
r_{\rm tot} = r_{\max }  - \Delta r \lambda 
\left( \frac{1}{\lambda_0}- \frac{1}{(\kappa_t \Delta r)}\right)
\label{eq:2ndGL}
\end{equation}
where $r_{\max }= 65.8\mu$M is a universal maximal ribosome concentration, $\Delta r = r_{\max}-r_{\min}=46.5\mu$M is the dynamic range of the active ribosome concentration and $\lambda_0$ is the bacterial growth rate in the absence of antibiotic. Eq. (\ref{eq:2ndGL}) states that the total ribosome content increases as growth rate decreases due to ribosome inhibition, but the slope of this increase depends on how fast the cells were growing before they were inhibited ({\em i.e.} on $\lambda_0$). Fast-growing cells, in a rich growth medium, increase their ribosome content proportionally less than do slow-growing cells, in a poor growth medium. Intuitively, fast-growing cells, which have a high ribosome content, already need to devote close-to-maximal protein production capacity to ribosome production so cannot increase ribosome synthesis further upon antibiotic challenge. In contrast, slow-growing cells, which have lower ribosome content, have excess protein production capacity that can be diverted to ribosome synthesis.  In our model the total ribosome concentration is given by $r_{\rm tot}=r_{\rm u}+r_{\rm b}$. For cells growing exponentially, the contents of the cell are in steady state, and thus the rate of ribosome synthesis must match the rate of ribosome removal 
by dilution: $s = \lambda r_{\rm{tot}}$. This leads to a quadratic expression for the synthesis rate $s$ as a function of $\lambda$:
\begin{equation}
s(\lambda)  = \lambda \left[ r_{\max }  - 
\lambda \Delta r\left( \frac{1}{\lambda_0}- \frac{1}{(\kappa_t \Delta r)}\right) \right].
\label{eq:SynthRate}
\end{equation}
Eqs. (\ref{eq:antibDyn})-(\ref{eq:rbDyn}) together with (\ref{eq:1stGL}) and (\ref{eq:SynthRate}) constitute a complete description of the model. \\

It is useful to comment on two points regarding the biological interpretation of this model. First, over times shorter than the bacterial generation time, the model describes  an exponentially growing bacterial cell.   To see this, we note that if the cell is growing exponentially, then its volume increases as $\dot{V} = \lambda{V}$. Assuming that molecules of a particular type are produced at a rate proportional to the volume, we obtain $\dot{N} = gV$ for the molecule number $N$, where $g$ is a production rate constant. The dynamics of the molecular concentration $n=N/V$ is then given by $\dot{n} = (1/V)\dot{N} - (N/V^2)\dot{V}=g-\lambda n$, as in Eqs (\ref{eq:antibDyn})-(\ref{eq:rbDyn}). Second, over times longer than the bacterial generation time, the model describes the behaviour of a lineage of cells. Bacterial cells undergo periodic division events; however, these events do not (on average) change the molecular concentrations, because both the molecule number and the cell volume are (on average) halved. Thus our model effectively describes an experiment in which we follow the dynamics of the molecular concentrations within an individual bacterial cell as it grows and divides, and in which, at each division event, we follow only one of the daughter cells. \\

Returning to the model equations (\ref{eq:antibDyn})-(\ref{eq:rbDyn}), (\ref{eq:1stGL}) and (\ref{eq:SynthRate}), it is useful to analyse the stationary points of the system, for which we set the  time derivatives in Eqs (\ref{eq:antibDyn})-(\ref{eq:rbDyn}) to zero: $\dot{a}=\dot{r_{\rm u}}=\dot{r_{\rm b}}=0$. Using Eq. (\ref{eq:1stGL}) to eliminate $r_{\rm u}$ in favour of $\lambda$,  then using eq. (\ref{eq:rbDyn}) to eliminate $r_{\rm b}$ in Eqs. (\ref{eq:antibDyn}) and (\ref{eq:ruDyn}) leads to two independent relations between $a$ and $\lambda$, which can be solved to eliminate $a$. This leads finally to  a cubic equation for the stationary points of the growth rate $\lambda$ \footnote{For details of this calculation see the Supplementary Material of Ref. \cite{greulich_2015}.}:
 
\[ 0 =  - \left( {\frac{\lambda }
{{\lambda _0 }}} \right)^3 \left(\frac{\lambda _0}{\lambda _0^{\rm{*}}}\right)^2\left[ {
\left( {1  + \frac{\kappa _t}{k_{\rm{on}}} } \right) } \right] +  
\left( {\frac{\lambda }
{{\lambda _0 }}} \right)^2 \left[ {\left( {1  + \frac{\kappa _t}{k_{\rm{on}}} } 
\right)\left(\frac{\lambda _0}{\lambda _0^{\rm{*}}}\right)^2  -\left( \frac{{P_{\rm out}  + 
k_{\rm off} }}{2\sqrt{P_{\rm out}k_{\rm off}}} \right)\left(\sqrt{\frac{\kappa _t 
}{k_{\rm{on}}}}\right)\left(\frac{\lambda _0}{\lambda_0^{\rm{*}}}\right) } \right] \]
\begin{equation}
   + \left( {\frac{\lambda }
{{\lambda _0 }}} \right)\left[{\left( \frac{{P_{\rm out}  + k_{\rm off} }}
{2\sqrt{P_{\rm out}k_{\rm off}}} \right)\left(\sqrt{\frac{\kappa _t}
{k_{\rm{on}}}}\right)\left(\frac{\lambda _0}{\lambda_0^{\rm{*}}}\right)  -  
\frac{{a_{\rm ex} }}
{2 {\rm IC}_{50}^*}\left(\frac{\lambda_0}{\lambda_0^{\rm{*}}}\right)  - 
\frac{1}{4} }\right] +  \frac{1}{4} , 
\label{eq:Sn2}
\end{equation}
where $K_{\rm D} =  k_{\rm off}/k_{\rm{on}}$ and we have defined the parameter combinations $\lambda_0^{\rm{*}} 
= 2 \sqrt{P_{\rm{out}}\kappa_tK_{\rm D}}$ and ${\rm IC}_{50}^* = \lambda_0^{\rm{*}}\Delta 
r/(2P_{\rm{in}})$. \\

\begin{figure}
\centering
\includegraphics[width=0.9\linewidth]{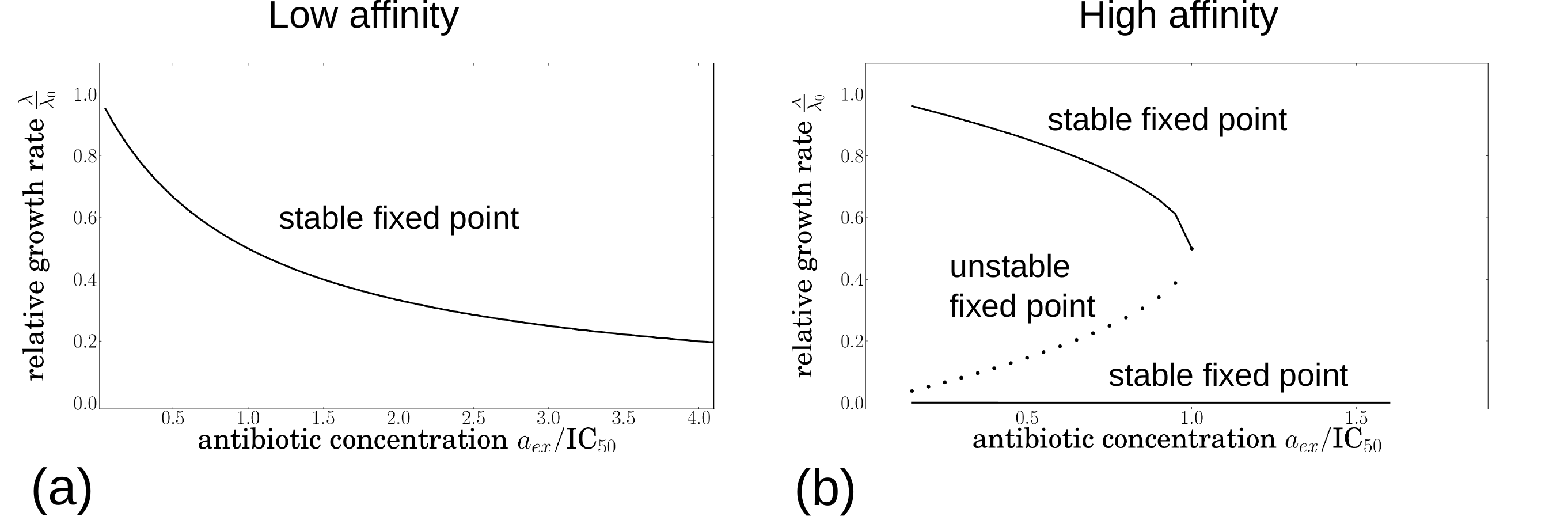}
\caption{Fixed points of Eq. (\ref{eq:Sn2}), plotted as a function of the external antibiotic concentration $a_{\rm ex}$, for the two parameter sets given in Table \ref{tab:1}, corresponding to antibiotics with large and small values of $\lambda_0^*$. Panel (a) shows results for a low-affinity antibiotic, with a large value of $\lambda_0^*$ (Table  \ref{tab:1}). Panel (b) shows  results for a high-affinity antibiotic, with a small value of $\lambda_0^*$ (Table  \ref{tab:1}). The fixed points were obtained numerically in python using sympy.solvers.solve. The external antibiotic concentration is scaled by the IC$_{50}$ value calcuated from Eq. (\ref{eq:IC50full}) and given in Table \ref{tab:1}.}
\label{fig:fixedpoints}
\end{figure}

Eq. (\ref{eq:Sn2}) is a key result of Ref. \cite{greulich_2015}. If the effective parameter $\lambda_0^{\rm{*}}$ is large, corresponding to reversible antibiotic transport and/or low-affinity binding (large values of $P_{\rm out}$ and/or  $K_D=k_{\rm off}/k_{\rm on}$), then Eq. (\ref{eq:Sn2}) has a single fixed point for any given value of $a_{\rm ex}$, as illustrated in Figure \ref{fig:fixedpoints} (a). This corresponds to a smoothly decreasing growth inhibition curve, as observed experimentally in Ref. \cite{greulich_2015} for the antibiotics chloramphenicol and tetracycline. In contrast, if the parameter $\lambda_0^{\rm{*}}$ is small, corresponding to irreversible transport and/or high-affinity binding (small values of $P_{\rm out}$ and/or $K_D=k_{\rm off}/k_{\rm on}$), then the model solution has 3 fixed points for values of $a_{\rm ex}$ below a critical threshold, which is a bifurcation point (Figure \ref{fig:fixedpoints} (b)). The upper and lower fixed points are stable sinks and the intermediate fixed point is an unstable saddle point of the dynamics.  For example, for the parameter set of Figure  \ref{fig:fixedpoints}(b), for $a_{\rm ex}=0.9\times$IC$_{50}$ (below the bifurcation point), the real parts of the eigenvalues of the Jacobean matrix at the three fixed points are $(-1.1\times10^4, -0.66, -0.90), (-5.7\times10^3, 0.89, -0.34), (-1.0\times10^6, -1.0\times10^{-3}, -2.7\times10^{-5})$. Thus, the first and third fixed points are stable (their eigenvalues have all negative real parts) while the second fixed point is unstable (it has an eigenvalue with a postive real part). For  $a_{\rm ex}=1.1\times$IC$_{50}$ (above the bifurcation point), the single fixed point has eigenvalues with all negative real parts $(-1.3\times 10^6, -1.0\times 10^{-2}, -2.2 \times 10^{-5})$ - {\em i.e.} it is stable.  The bifurcation points of the model are discussed in more detail in Appendix A, where it is also argued that the value $a^*_{\rm ex} = \Delta r \lambda_0/(4 P_{\rm in})$ is a good approximation for the (upper) bifurcation point.\\

From a practical point of view, for small values of the external antibiotic concentration, we expect to observe little inhibition of bacterial growth, corresponding to the upper fixed point of the dynamics. However, for antibiotic concentrations above the bifurcation point, we expect to see drastic growth inhibition, corresponding to the single (lower) fixed point. This implies a steep, threshold-like growth inhibition curve, as observed experimentally in Ref. \cite{greulich_2015} for the aminoglycosides streptomycin and kanamycin \footnote{It is important to recognise that streptomycin and kanamycin have other physiological effects, not included in our model, such as production of misfolded protein which may affect membrane permeability \cite{review_streptom_irr}).}. Eq. (\ref{eq:Sn2}) can also be used to derive a simple expression for the dependence of the IC$_{\rm 50}$ on $\lambda_0$:
\begin{equation}
\frac{\rm{IC}_{50}}{{\rm IC}_{50}^*} = 
\frac{1}{2}\left[\left(1+\frac{\kappa_t}{k_{\rm on}}\right)\left(\frac{\lambda_0}{\lambda_0^{\rm{*}}}\right)+ 
\left(\frac{\lambda_0^{\rm{*}}}{\lambda_0}\right)+\frac{\left(P_{\rm out}+k_{\rm off}\right)}{\sqrt{P_{\rm out}k_{\rm off}}}\sqrt{\frac{\kappa_t}{k_{\rm on}}}\right].
\label{eq:IC50full}
\end{equation}

Eq. (\ref{eq:IC50full}) predicts that antibiotic efficacy will increase with nutrient richness (IC$_{\rm 50}$ decreases with $\lambda_0$) when $\lambda_0^{\rm{*}}$ is large, but that  efficacy will decrease with nutrient richness (IC$_{\rm 50}$ increases with $\lambda_0$) when $\lambda_0^{\rm{*}}$ is small. These predictions, which are in agreement with experimental data, were discussed in detail in Ref. \cite{greulich_2015}. \\

\section*{Model predictions for dynamical response to antibiotic}

In a clinical context, antibiotic concentrations vary in time. In this paper, we explore the predictions of the model defined by Eqs (\ref{eq:antibDyn})-(\ref{eq:rbDyn}), (\ref{eq:1stGL}) and (\ref{eq:SynthRate}), for the response of bacterial growth rate to a time-dependent exposure to antibiotic -- {\em i.e.} we explore the dynamics $a(t)$, $r_{\rm u}(t)$ and $r_{\rm b}(t)$ for a time-varying external antibiotic concentration $a_{\rm ex}(t)$. In most cases (with some exceptions that we discuss below), these equations are not amenable to analytical solution in the time-varying case. We therefore obtain predictions  by integrating the model equations numerically \footnote{All numerical solutions of the system of ordinary differential equations described in this paper were carried out in python using scipy.integrate.odeint.}. We compare results for two sets of parameters, representing antibiotics which are bound and transported with ``low affinity'' (high values of $P_{\rm out}/P_{\rm in}$ and $k_{\rm off}/k_{\rm on}$) and with ``high affinity'' (low values of $P_{\rm out}/P_{\rm in}$ and $k_{\rm off}/k_{\rm on}$). These parameters, which are chosen to be within the range of literature values for tetracycline and streptomycin respectively \footnote{Literature values for kinetic parameters for tetracycline, chloramphenicol, streptomycin and kanamycin are reviewed in the Supplementary Material of Ref. \cite{greulich_2015}.}, are listed in Table \ref{tab:1}.\\

 \begin{table}
\caption{Parameter values used in this study to model low and high-affinity ribosome-targeting antibiotics. These values are chosen to be within the range of the literature values collated in Ref. \cite{greulich_2015}. The universal parameters  are $\kappa_t=6.1\times 10^{-2} \mu$M\,h$^{-1}$, $r_{\rm min}=19.3\mu$M and $r_{\rm max}=65.8\mu$M \cite{scott_science,greulich_2015}. Except where stated otherwise, we have assumed an antibiotic-free growth rate $\lambda_0$ of 1h$^{-1}$.}
    \begin{tabular}{c c c}
    Parameter & Value for low-affinity antibiotic & Value for high-affinity antibiotic  \\
   \hline
    $P_{\rm in}$ & 2000h$^{-1}$ & 1h$^{-1}$\\
    $P_{\rm out}$ & 100h$^{-1}$  & 0.01h$^{-1}$\\
    $k_{\rm on}$ & 1000$\mu$M$^{-1}$h$^{-1}$ & 1000$\mu$M$^{-1}$h$^{-1}$\\
    $k_{\rm off}$ &  $10^5$h$^{-1}$ & 10h$^{-1}$\\\hline
$\lambda_0^* = 2\sqrt{P_{\rm out}\kappa_tk_{\rm off}/k_{\rm on}}$ & 49.4h$^{-1}$ & 0.00493h$^{-1}$\\
IC$_{50}^*=\lambda_0^*(r_{\rm max}-r_{\rm min})/(2P_{\rm in})$ & 0.574$\mu$M& 0.115$\mu$M\\
Predicted IC$_{50}$ (from Eq. (\ref{eq:IC50full})) & 1.43$\mu$M & 11.64$\mu$M
\end{tabular}
\label{tab:1}
\end{table}   

It is important to note that the growth laws which we use in our model, Eqs.  (\ref{eq:1stGL}) and (\ref{eq:SynthRate}), are derived from experimental measurements on exponentially growing bacteria, for which all intracellular concentrations are in steady state. In using 
 these constraints to make predictions for dynamical trajectories we assume that the cell adjusts its rates of growth and ribosome synthesis rapidly in response to changing external conditions, in comparison to the rate at which the external conditions vary. It is known that the ribosome synthesis rate can adjust within minutes to changes in nutrient conditions \cite{schaechter1958}. A typical timescale for synthesis of a protein molecule is $\sim$1 minute ($\sim$ 1000 amino acids polymerised at a translation rate of $\sim$ 20 amino acids per second \cite{bremer_dennis}), while a conservative estimate for the timescale for synthesis of a ribosome is $\sim$6 minutes ($\sim$ 7500 amino acids in the entire ribosomal complex, produced at $\sim$ 20 amino acids per second \cite{bremer_dennis}). Although in this paper we consider for simplicity step-like changes in antibiotic concentration, the timescale over which antibiotic concentration builds up in the body after an oral dose is $\sim$30 minutes, with a slower decay time due to excretion \cite{greenwood}. The use of the steady-state constraints (\ref{eq:1stGL}) and (\ref{eq:SynthRate}) therefore seems reasonable.

\subsection*{Response to a step increase in antibiotic concentration}

To analyse the dynamical behaviour of the model, we first consider the response to a sudden, step-like increase in antibiotic concentration, from zero to a fixed value: $a_{\rm ex}(t)=0$ for $t<t_0$ and $a_{\rm ex}(t)=a_{\rm ex}^{\rm final}$ for $t>t_0$. Although a step-increase in antibiotic concentration is unlikely in a clinical context, it is achievable in laboratory experiments (e.g. using a morbidostat \cite{toprak_morbidostat} or microfluidic flow device \cite{lin_2016}).

\subsubsection*{Low-affinity antibiotic}



\begin{figure}[h!]
\centering
\includegraphics[width=0.9\linewidth,clip=true]{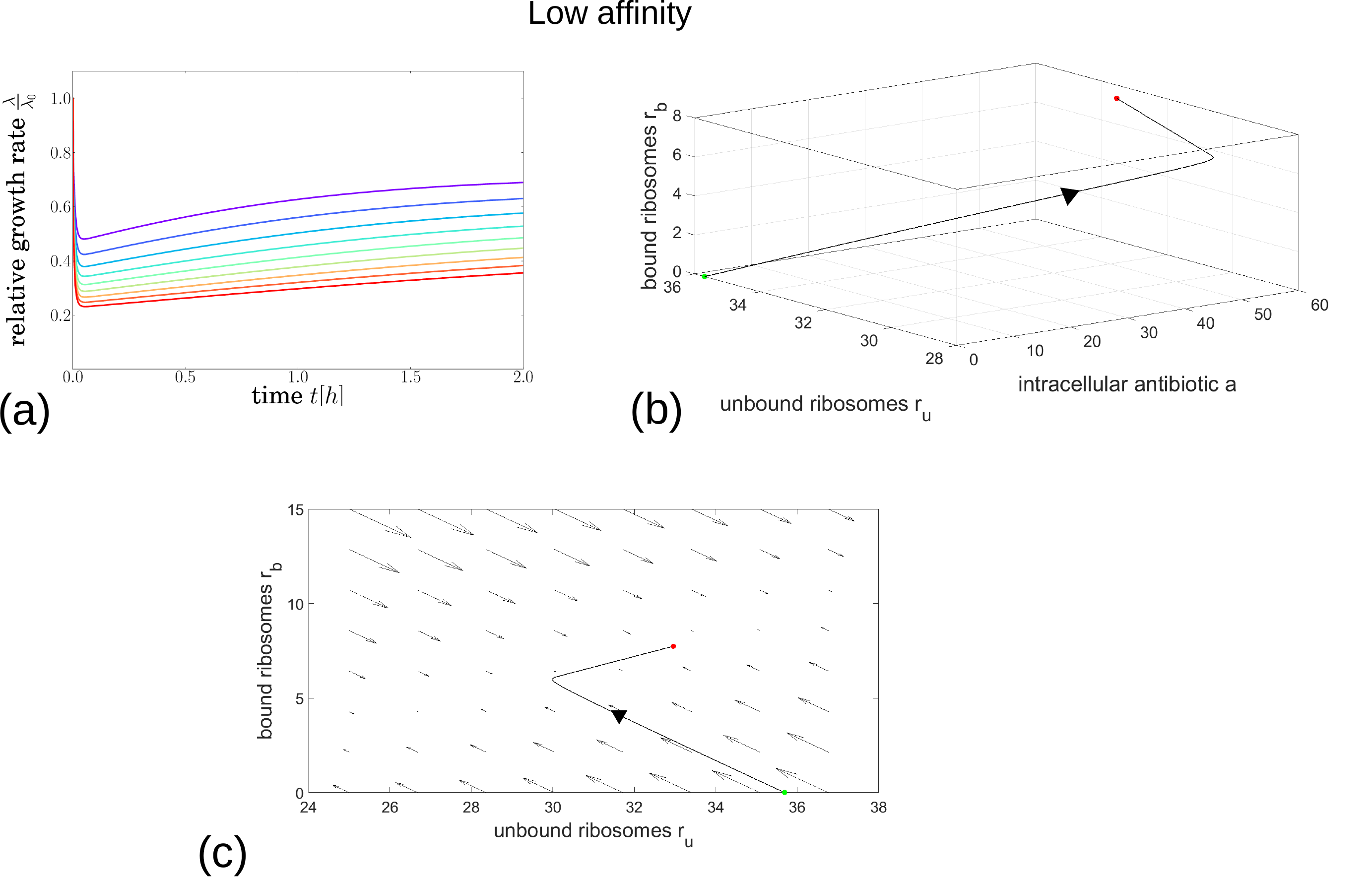}
\caption{Dynamical trajectories showing growth inhibition after  a step increase in antibiotic concentration, for the low-affinity parameter set (with parameter values as in Table \ref{tab:1}). (a): Relative growth rate $\lambda(t)/\lambda_0$ as a function of time after the step increase in antibiotic concentration. The final antibiotic concentration $a_{\rm ex}^{\rm final}$ is indicated by the line colour, ranging from $0.4\times$IC$_{\rm 50}$ (purple) to $1.3\times$IC$_{\rm 50}$ (red), in steps of $0.1\times$IC$_{\rm 50}$. (b): Trajectory in the 3-dimensional space of variables $a$, $r_{\rm u}$ and $r_{\rm b}$, for the case of a step increase to $a_{\rm ex}^{\rm final}=2\times$IC$_{50}$. The initial and final system states are shown by the green and red points respectively. (c): Flow diagram showing the direction and magnitude of the flow field $\dot{r_{\rm u}},\dot{r_{\rm b}}$ for a fixed value of $a$ corresponding to the final point of the trajectory in panel (b), and for $a_{\rm ex}=2\times$IC$_{50}$. The trajectory from (b) is also shown here, projected onto the $r_{\rm u}, r_{\rm b}$ plane.}
\label{fig:trajectories_rev}
\end{figure}

Figure \ref{fig:trajectories_rev} explores the dynamical response of the model to a step increase in antibiotic concentration, for the low-affinity antibiotic. The model predicts a strikingly non-monotonic response of the bacterial growth rate $\lambda(t)/\lambda_0$, as shown in Figure  \ref{fig:trajectories_rev}(a): we observe an initial rapid decrease in growth rate, followed by a slower recovery to a steady-state value that depends on the antibiotic concentration $a_{\rm ex}^{\rm final}$. This steady-state value corresponds to the fixed point of the model dynamics (Figure \ref{fig:fixedpoints}(a)). The origin of this non-monotonic response can be understood by plotting the trajectory of the model in the 3d space of its variables $a$, $r_{\rm u}$ and $r_{\rm b}$, as in Figure  \ref{fig:trajectories_rev}(b). Following the increase in $a_{\rm ex}$, the intracellular antibiotic concentration $a$ rapidly increases, accompanied by a decrease in the concentration of unbound ribosomes and an increase in the concentration of bound ribosomes $r_{\rm b}$. These changes are driven by the rapid dynamics of antibiotic transport and binding/unbinding. The later, much slower, recovery of the growth rate observed  in Figure \ref{fig:trajectories_rev}(a) corresponds to an increase in both $r_{\rm u}$ and $r_{\rm b}$ in the trajectory of \ref{fig:trajectories_rev}(b) and is associated with the slower dynamics of ribosome synthesis in response to the antibiotic challenge. Thus, the non-monotonic response of the growth rate predicted by the model is due to the initial, rapid processes of transport and binding, followed by a slower partial recovery due to increased ribosome synthesis. \\

The dynamics of the model can also be illustrated in the form of a flow diagram, as in Figure \ref{fig:trajectories_rev}(c). Here, the arrows show the direction of the flow field $\dot{r_{\rm u}},\dot{r_{\rm b}}$ for a fixed value of $a$. The trajectory of Figure \ref{fig:trajectories_rev}(b) is shown projected onto this plane. This diagram illustrates clearly the separation of timescales between transport and binding, which produce a strong flow field towards the centre of the diagram, and ribosome synthesis, which is responsible for the slower dynamics as the system approaches the stable fixed point.

\subsubsection*{High-affinity antibiotic}

\begin{figure}[h!]
\centering
\includegraphics[width=0.9\linewidth,clip=true]{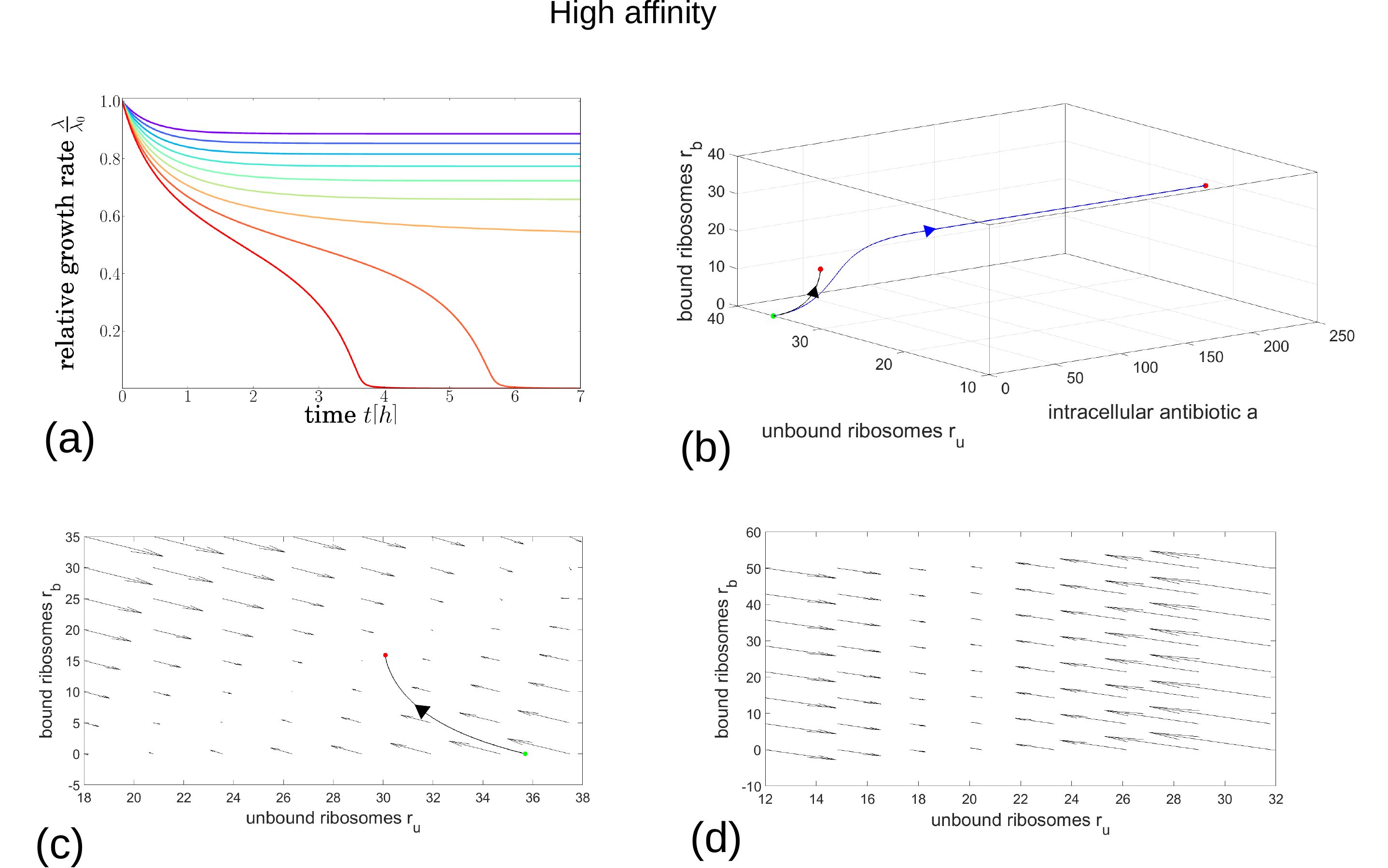}
\caption{Dynamical trajectories showing growth inhibition after  a step increase in antibiotic concentration, for the high-affinity parameter set (with parameter values as in Table \ref{tab:1}). (a): Relative growth rate $\lambda(t)/\lambda_0$ as a function of time after the step increase in antibiotic concentration. The final antibiotic concentration $a_{\rm ex}^{\rm final}$ is indicated by the line colour, ranging from $0.4\times$IC$_{\rm 50}$ (purple) to $1.3\times$IC$_{\rm 50}$ (red), in steps of $0.1\times$IC$_{\rm 50}$. (b): Trajectory in the 3-dimensional space of variables $a$, $r_{\rm u}$ and $r_{\rm b}$, for the case of a step increase to $a_{\rm ex}^{\rm final}=0.9\times$IC$_{50}$ (black line) and $a_{\rm ex}^{\rm final}=1.1\times$IC$_{50}$ (blue line). The initial and final system states are shown by the green and red points respectively. (c): Flow diagram showing the direction and magnitude of the flow field $\dot{r_{\rm u}},\dot{r_{\rm b}}$ for a fixed value of $a$ corresponding to the final point of the black trajectory in panel (b), and for $a_{\rm ex}=0.9\times$IC$_{50}$. The trajectory from (b) is also shown here, projected onto the $r_{\rm u}, r_{\rm b}$ plane. (d): Flow diagram showing the flow field $\dot{r_{\rm u}},\dot{r_{\rm b}}$ for a fixed value of $a=200\mu$M  and for $a_{\rm ex}=0.9\times$IC$_{50}$.}
\label{fig:trajectories_irrev}
\end{figure}

The model predictions are strikingly different for the high-affinity antibiotic (Figure  \ref{fig:trajectories_irrev}). The bacterial growth rate (Figure  \ref{fig:trajectories_irrev} (a)) is predicted to decrease smoothly and monotonically for low antibiotic concentrations $a_{\rm ex}^{\rm final}$, as it approaches the upper fixed point in Figure \ref{fig:fixedpoints}(b).  However, for antibiotic concentrations  $a_{\rm ex}^{\rm final}$ that are above the bifurcation point in Figure \ref{fig:fixedpoints}(b) the model predicts instead a decline in growth rate to a state in which there is essentially no growth. The timescale of this approach to the non-growing steady state can be very long (of the order of days) for antibiotic concentrations close to the bifurcation point. \\

Plotting the dynamics of the model in the 3d space of its variables, as in Figure \ref{fig:trajectories_irrev}(b), illustrates the very different nature of its trajectories for values of $a_{\rm ex}^{\rm final}$ below and above the bifurcation point. If $a_{\rm ex}^{\rm final}$ is below the bifurcation point, as for the black trajectory (for $a_{\rm ex}^{\rm final}=0.9\times$IC$_{50}$), the dynamics approaches a fixed point which is close to the initial state, and in which the intracellular antibiotic concentration is small. This corresponds to the upper stable fixed point in Figure \ref{fig:fixedpoints}(b). However, if $a_{\rm ex}^{\rm final}$ is above the bifurcation point, as for the blue trajectory (for $a_{\rm ex}^{\rm final}=1.1\times$IC$_{50}$), the dynamics instead approaches a very different state, with a far higher intracellular antibiotic concentration and with $r_{\rm u}$ close to $r_{\rm min}$: this corresponds to the lower fixed point in Figure \ref{fig:fixedpoints}(b), with essentially no growth. \\

The flow diagrams of Figure \ref{fig:trajectories_irrev}(c) and (d) illustrate the two stable fixed points of the model dynamics for values of  $a_{\rm ex}$ below the bifurcation point. These diagrams show the  flow field $\dot{r_{\rm u}},\dot{r_{\rm b}}$ for $a_{\rm ex}=0.9\times$IC$_{50}$, for two different values of the intracellular antibiotic concentration $a$. In Figure \ref{fig:trajectories_irrev}(c), the flow field is shown for $a=0.016\mu$M, which corresponds to the final point of the black trajectory in Figure \ref{fig:trajectories_irrev}(b) (this trajectory is also shown, projected onto the  $a=0.016\mu$M plane). The model has a stable fixed point for values of $r_{\rm u}$ and $r_{\rm b}$ which are close to the starting point of the trajectory ({\em i.e.} the system state in the absence of antibiotic). The second stable fixed point is evident in Figure \ref{fig:trajectories_irrev}(d), which shows the flow field for $a=200\mu$M. This fixed point occurs at a much smaller value of $r_{\rm u}$ ($r_{\rm u}\approx r_{\rm min}$) and a higher value of $r_{\rm b}$. Interestingly, as for the low-affinity case (Figure  \ref{fig:trajectories_rev}(c)), the flow diagrams of Figure \ref{fig:trajectories_irrev} show a separation of timescales between the rapid dynamics of transport and binding and the slower dynamics of ribosome synthesis. However the separation is less extreme than for the low-affinity case (since $P_{\rm in}$, $P_{\rm out}$ and $k_{\rm off}$ are all smaller) -- this may explain why the approach to the stable state is monotonic rather than non-monotonic for our high-affinity parameter set.

\subsection*{Time to full inhibition}

\begin{figure}
\centering
\includegraphics[width=0.9\linewidth]{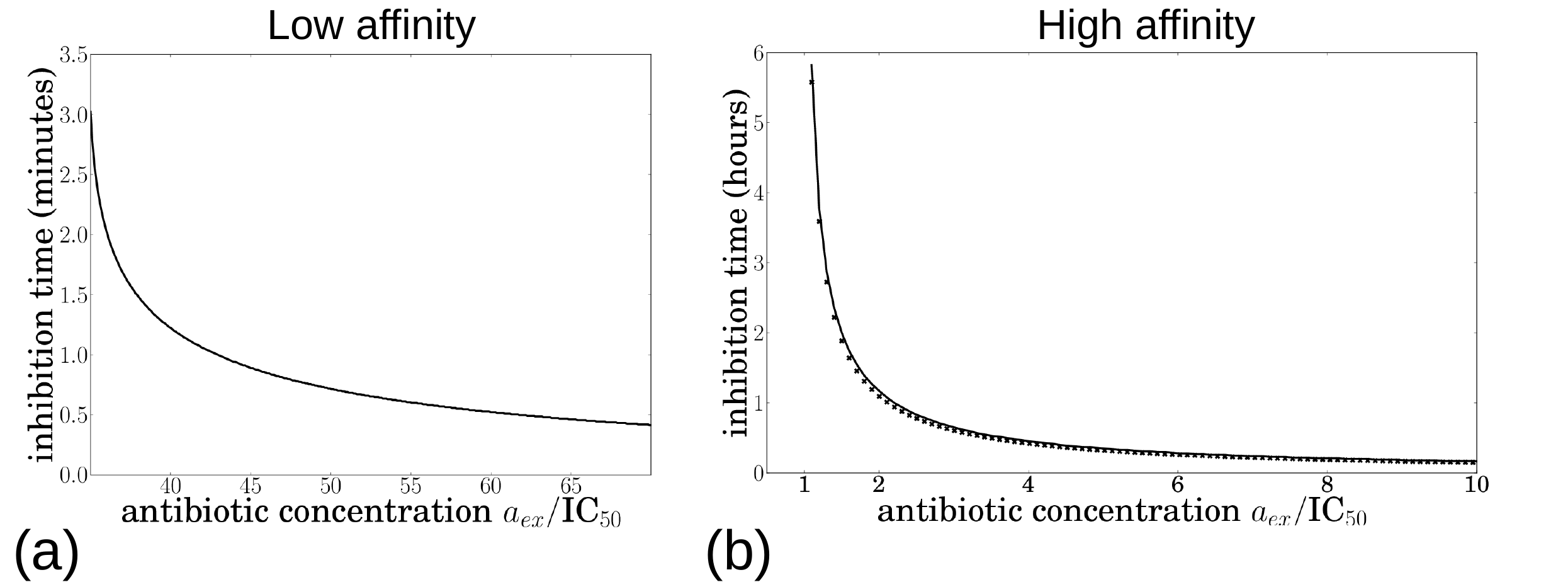}
\caption{Model predictions for the time required to reach 99\% growth inhibition $\lambda/\lambda_0=0.01$, following a step increase in antibiotic concentration. The inhibition time is plotted as a function of the external antibiotic concentration (note that the time is measured in minutes in (a) but in hours in (b)). Panel (a) shows results for the low-affinity antibiotic; panel (b) shows results for the high-affinity antibiotic (using parameters as in Table \ref{tab:1}); here the numerical solution of the model is shown by the solid line while the crosses show the analytical prediction based on the adiabatic approximation, described in Appendix B. For antibiotic concentrations lower than those plotted here, the dynamical trajectory of $\lambda/\lambda_0$ always stays above 0.01.}
\label{fig:fullinhibition}
\end{figure}

From a practical point of view, one may be interested in the time taken to achieve full growth inhibition following a high-concentration antibiotic dose. Because the variables in our model are continuous, the growth rate never completely reaches zero, but as a proxy for full inhibition we measure the time taken to achieve 99\% inhibition, {\em i.e.} to reduce the growth rate to 1\% of its antibiotic-free value: $\lambda/\lambda_0= 0.01$. Although this is an arbitrary threshold, we note that in a clinical situation a drastic reduction in bacterial population density is expected to lead to elimination of an infection, due to the action of the immune system \cite{pankey}. In our model, 99\% inhibition only occurs for high concentrations of antibiotic; for lower concentrations the system instead reaches a steady state with a growth rate greater than $0.01\times\lambda_0$ (as shown in Figures \ref{fig:trajectories_rev}(a) and \ref{fig:trajectories_irrev}(a)).\\

Figure \ref{fig:fullinhibition} shows the time to reach $\lambda=0.01\times\lambda_0$, as a function of the antibiotic concentration, for the low-affinity and high-affinity antibiotics. As expected, higher antibiotic concentrations lead to more rapid growth inhibition. For the low-affinity antibiotic (Figure  \ref{fig:fullinhibition}(a)), a high concentration is needed to achieve 99\% inhibition ($\sim 35\times$IC$_{50}$), but for these concentrations, 99\% inhibition is achieved very rapidly, on a timescale of minutes. The inhibition time decreases smoothly as the antibiotic concentration increases. This is consistent with the inhibition trajectories shown in Figure \ref{fig:trajectories_rev}(a), which show a rapid initial inhibition of growth, followed by a slower partial recovery to the steady state. For the high-affinity antibiotic (Figure  \ref{fig:fullinhibition}(b)), 99\% inhibition is achieved for much lower concentrations of antibiotic, just above the IC$_{50}$, but the timescale for inhibition is much longer, of the order of hours for concentrations close to the IC$_{50}$. This is consistent with the inhibition trajectories of Figure \ref{fig:trajectories_irrev}(a), which show very long timescales for inhibition for external antibiotic concentrations close to the bifurcation point of the model dynamics. As we discuss in Appendix B (and illustrate in Figure \ref{fig:philip}), this very slow inhibition occurs because of a ``bottleneck'' effect, in which dynamical trajectories slow down as they pass close to the location where the two fixed points have merged. \\

For the high-affinity antibiotic, it is possible to obtain an analytical prediction for the time to achieve 99\% inhibition, by making an adiabatic approximation for the dynamics of the intracellular antibiotic concentration $a(t)$. This calculation is presented in detail in Appendix B; briefly, we assume that the dynamics of $a(t)$ are fast compared to those of the other variables, and set $\dot{a}=0$ in Eqs (\ref{eq:antibDyn})-(\ref{eq:rbDyn}). This reduces the model to a set of dynamical equations for $r_{\rm u}$  and $r_{\rm b}$, and setting $k_{\rm off}=0$ (for an irreversible antibiotic) decouples these equations, allowing one to solve for  $r_{\rm u}$, and hence for the growth rate $\lambda(t)$ via the constraint (\ref{eq:1stGL}). Figure  \ref{fig:fullinhibition}(b) shows that the resulting analytical prediction for the inhibition time (crosses) is in good agreement with the numerical results (solid line).



\subsection*{Response to a step pulse of antibiotic}

\begin{figure}
\centering
\includegraphics[width=0.9\linewidth]{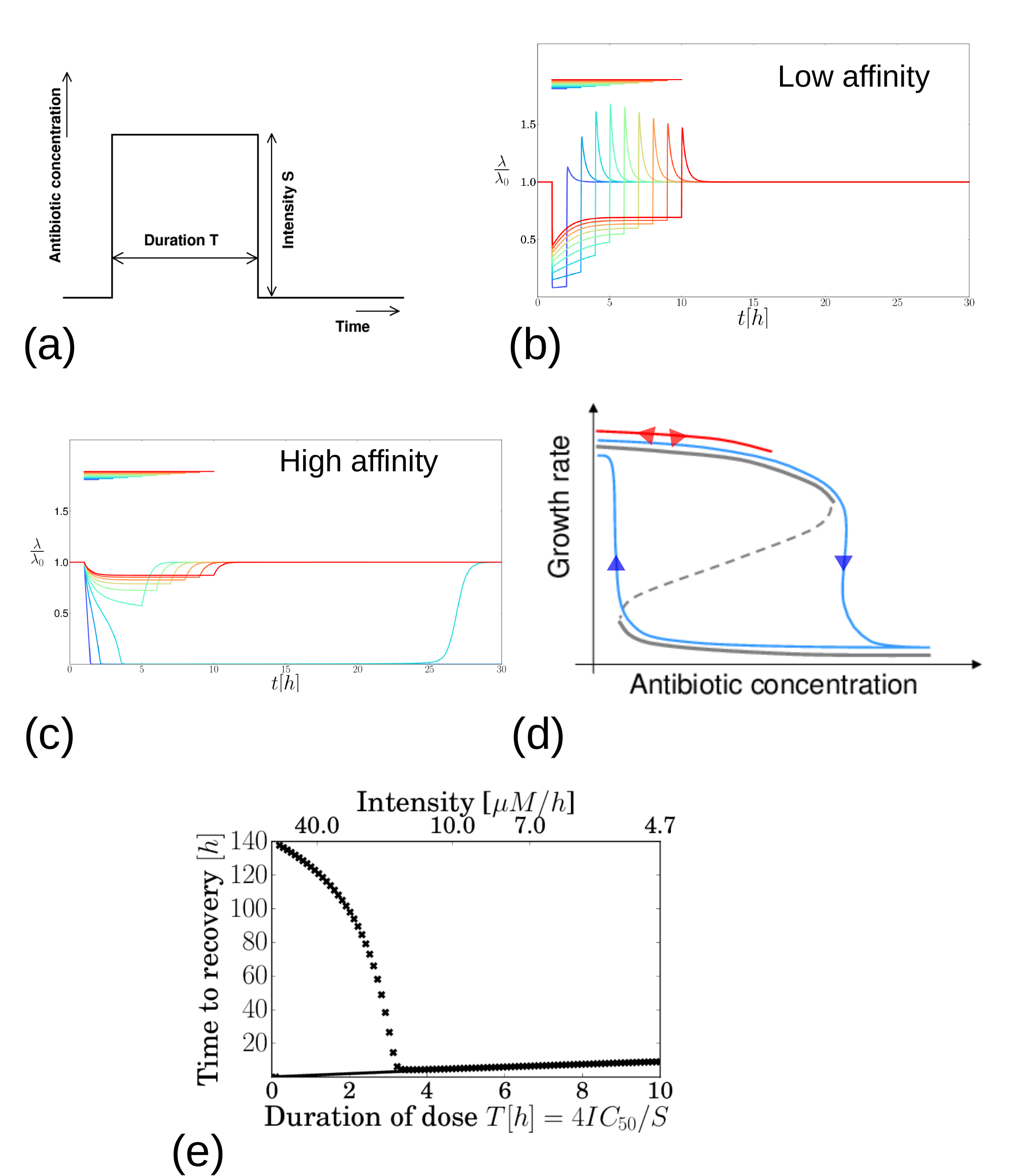}
\caption{Growth inhibition in response to a transient step-like dose of antibiotic. (a): Illustration of the dosing protocol: antibiotic concentration is switched suddenly to a value $S$ at the start of the dose, and is switched back to zero after a time $T$. The total dose $S\times T$ is fixed at $4\times$IC$_{\rm 50}$. (b) Growth-rate trajectories for the low-affinity antibiotic (with parameters as in Table \ref{tab:1}). The coloured lines represent doses of different duration and intensity (keeping the total dose fixed at  $4\times$IC$_{\rm 50}$). The colour bars show the duration of the dose. (c) As in panel (b), but for the high-affinity antibiotic (with parameters as in Table \ref{tab:1}). (d) Schematic illustration of hysteresis in the model for the high-affinity antibiotic. The response to a low-intensity pulse of antibiotic is shown by the red line: upon addition of antibiotic the system tracks the upper stable fixed point and reverses its trajectory when the antibiotic is removed. The response to a high-intensity pulse is shown by the blue lines: upon addition of antibiotic the system transitions to the lower stable fixed point, and it tracks the lower fixed point when antibiotic is removed. (e) Time taken to recover from a step dose of antibiotic, as a function of the duration of the dose.  The solid line shows results for the low-affinity antibiotic, the symbols show results for the high-affinity antibiotic. Here, ``recovery'' is defined to mean that the growth rate $\lambda$ returns to a value $0.9\times\lambda_0$, having previously fallen below this threshold. The recovery time is defined as the total time during which the growth rate is suppressed below $0.9\times\lambda_0$.
}
\label{fig:transientstep}
\end{figure}

In a clinical situation, antibiotic treatment has a finite duration. The local antibiotic concentration at the infection site increases after a dose is given and later decreases due to removal of the antibiotic from the body (the pharmocokinetic curve \cite{greenwood}). To mimic such a scenario, we  investigate the response of the model to pulses of antibiotic of finite duration.  We obtain predictions for the dynamics of the bacterial growth rate during and after the dose, for low- and high-affinity antibiotics, and for different dose durations and intensities. For simplicity, we first consider a step pulse of antibiotic of intensity $S$  that is maintained for a fixed time $T$, as illustrated in Figure \ref{fig:transientstep}(a); later we also consider a continuously varying dose function.  We compare results for a fixed total antibiotic dose (duration $\times$ intensity)  -- {\em i.e.} we compare the effect of a short, high-intensity dose with that of a long, low-intensity dose. Specifically, we fix $S \times T=4\times$IC$_{50}$. Although this choice is arbitrary, we find qualitatively similar results for other values of the total dose.\\

\subsubsection*{Low-affinity antibiotic: growth-rate overshoot following antibiotic removal}

Figure \ref{fig:transientstep}(b) shows the model predictions for the bacterial growth rate, during and after a step-like pulse of a low-affinity antibiotic. The colours  indicate doses of varying duration (as shown by the bars). During the dose, bacterial growth is suppressed, to a degree that depends on the intensity of the dose (the short, high intensity dose shown by the blue line causes a greater degree of growth inhibition than the long, low intensity dose shown by the red line). Interestingly, the model also predicts a ``growth rate overshoot'' phenomenon: a peak in $\lambda(t)$ after the antibiotic dose ends, implying a transient increase in growth rate {\em above} the antibiotic-free steady-state value $\lambda_0$. This overshoot is predicted to be significant, with the growth rate increasing to more than $1.5\times \lambda_0$ for intermediate dose durations. The overshoot occurs because, in our model, ribosome synthesis is upregulated during exposure to the antibiotic ($s$ is larger, according to Eq. (\ref{eq:SynthRate})), such that the total ribosome concentration is higher than it would be in the absence of antibiotic. Once the external antibiotic is removed, intracellular antibiotic dissociates rapidly from bound ribosomes, since $k_{\rm off} \gg \lambda_0$, so that the free ribosome pool becomes transiently larger than it would have been in the absence of antibiotic. In our model, this produces a transient increase in growth rate. The magnitude of the effect is greatest at intermediate antibiotic dose duration; this is because for very short antibiotic pulses, the bacterium does not have time to increase its ribosome pool significantly before the pulse ends, while for very long pulses the antibiotic concentration (pulse intensity) is not high enough to produce a significant upregulation of ribosome concentration. Consistent with this explanation, when we repeat our simulations keeping the dose intensity fixed ({\em i.e.} increasing total dose as the duration increases), we find that the maximal overshoot occurs for the longest dose duration. \\

Upregulation of ribosome synthesis upon exposure to antibiotic is a growth medium-dependent phenomenon: for bacteria growing in a poor medium (with a small drug-free growth rate $\lambda_0$), the relative increase of the ribosome synthesis rate is larger than for bacteria growing on rich medium (with a large $\lambda_0$) \cite{scott_science}. This growth medium dependence is captured by the $\lambda_0$-dependence of the synthesis rate $s$ in our model (Eq. (\ref{eq:SynthRate})). We therefore expect that the magnitude of the growth-rate overshoot predicted by the model will be medium-dependent, with a larger overshoot for small values of $\lambda_0$, {\em i.e.} for bacteria growing on poor medium, because these bacteria upregulate ribosome synthesis more strongly during the antibiotic pulse and therefore have a greater excess of ribosomes after the pulse. Indeed, upon repeating our calculations for a range of values of $\lambda_0$, we observe a strong $\lambda_0$-dependence of the magnitude of the overshoot. For example, for a dose of duration $\sigma=7$h (orange curve in Fig. \ref{fig:transientstep}(b)), the growth rate at the peak of the overshoot is predicted to be $\lambda/\lambda_0=2.3,1.7,1.3$, for drug-free growth rates of $\lambda_0=0.5,1.0,1.5$h$^{-1}$ respectively. \\

\subsubsection*{High-affinity antibiotic: post-antibiotic growth suppression and hysteresis}

Figure \ref{fig:transientstep}(c) shows equivalent  predictions for the growth-rate response to a step pulse of high-affinity antibiotic. Here we observe a different phenomenon: the qualitative nature of the response is highly intensity-dependent. For long-duration, low-intensity doses  the growth rate is suppressed during the dose but recovers quickly when the antibiotic is removed (red-green curves in Figure \ref{fig:transientstep}(c)). However, for shorter, high intensity doses, the model shows a significant post-antibiotic effect: the growth rate decreases almost to zero during the dose and does not recover until many hours after the dose has ended  (blue curves in Figure \ref{fig:transientstep}(c)).  This phenomenon arises from hysteresis in the model. When antibiotic is added, the fixed points of the model move along the $a_{\rm ex}$ axis in Figure \ref{fig:fixedpoints}(b). As illustrated schematically in Figure \ref{fig:transientstep}(d), for a low-intensity antibiotic dose, the system tracks the upper stable fixed point and reverses its trajectory when the antibiotic is removed (red line in Figure \ref{fig:transientstep}(d)). This corresponds to the red-green trajectories in Figure \ref{fig:transientstep}(c). However, for a high-intensity antibiotic dose, the system is pushed past the bifurcation point in Figure \ref{fig:fixedpoints}(b), forcing it to transition to the  lower stable fixed point in which the growth rate is close to zero. When the antibiotic is removed, the system moves back along the lower line of fixed points, before eventually transitioning back to the upper fixed point (blue lines in Figure \ref{fig:transientstep}(d)). The timescale over which this eventual recovery happens is controlled by the antibiotic-ribosome dissociation rate constant $k_{\rm off}$, which is small for the high-affinity antibiotic. Although we always see eventual recovery of the bacterial growth rate in our simulations, in a clinical setting we expect that other factors, such as immune response, would lead to elimination of the infection \cite{pankey}. \\


\subsubsection*{Optimal dosing strategy differs for low and high-affinity antibiotics}

In a clinical setting, antibiotic dosing protocols target different features of the pharmacokinetic curve: some are designed to maximise the peak antibiotic concentration, while others aim to maximise the time the concentration is above a threshold, or the area of the curve above the threshold \cite{greenwood,craig}. Although our simulated step-like dosing protocol (Figure \ref{fig:transientstep}(a)) is simplistic, we do see clear differences in optimal dosing strategy for low-affinity and high-affinity antibiotics. These differences are illustrated in Figure \ref{fig:transientstep}(e), where we plot the time required to recover from a step-like antibiotic dose, as a function of the duration of the dose (and hence its inverse intensity, as shown on the upper horizontal axis). Here, we define time to recovery as the time taken for the growth rate $\lambda$ to recover to 90\% of its antibiotic-free steady state value $\lambda_0$, having previously falled to below this value. For the low-affinity antibiotic (Figure \ref{fig:transientstep}(e), solid line), the recovery time is proportional to dose duration: this is consistent with the growth inhibition trajectory (Figure \ref{fig:transientstep}(b)), in which growth is suppressed during the dose and recovers rapidly afterwards. Therefore, for ribosome-targeting antibiotics which bind with low affinity and/or are transported reversibly, the model suggests that an optimal protocol would maximise the time over which the dose is maintained above a threshold. In contrast, for the high-affinity antibiotic (Figure \ref{fig:transientstep}(e), symbols), the model predicts that the recovery time increases dramatically, to many times longer than the dose, when the dose intensity exceeds a well-defined threshold ({\em i.e.} for shorter dose durations in our simulations). This is also consistent with the growth inhibition trajectories of \ref{fig:transientstep}(c). Thus our model suggests that for ribosome-targeting antibiotics which bind with high affinity and/or are transported irreversibly, it may be more important to maximise the peak concentration of the pharmacokinetic curve than the duration of the dose. This prediction is consistent with the observation that aminoglycoside antibiotics can show significant post-antibiotic effects \cite{bundtzen_1981,isaksson_1988,mackenzie_1993,vogelman,craig}.

\subsection*{Response to a smooth pulse of antibiotic}

\begin{figure}
\centering
\includegraphics[width=0.9\linewidth]{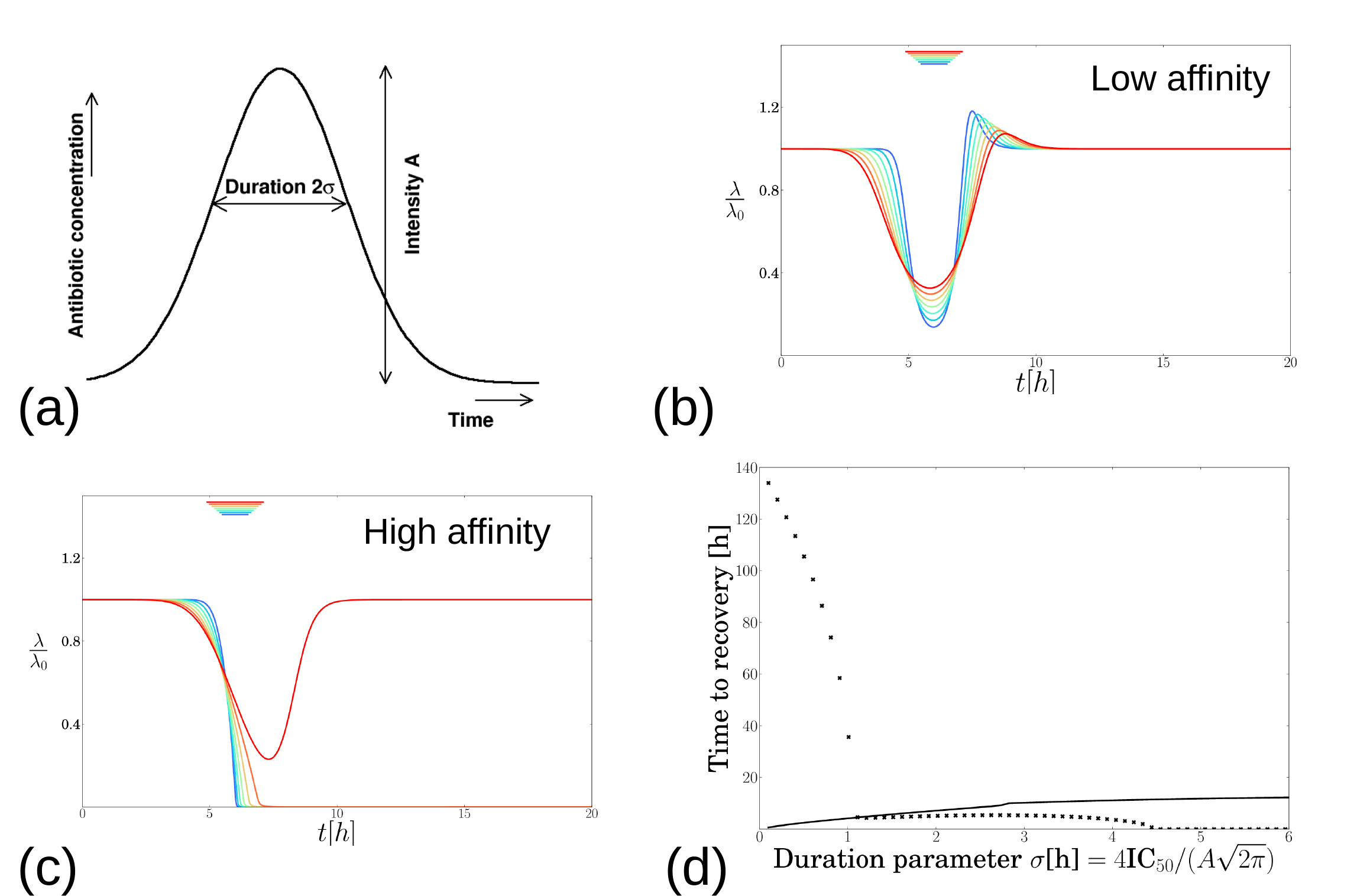}
\caption{Growth inhibition in response to a transient dose of antibiotic with a Gaussian profile. (a): Illustration of the dosing protocol. The total dose, which is given by $\sqrt{2\pi} A \sigma$, is fixed at $4\times$IC$_{50}$. (b) Results for the low-affinity antibiotic, with parameters as in Table \ref{tab:1}. The coloured lines represent doses of different duration and intensity, from $\sigma =0.4$h (blue) to $\sigma=1.1$h (red), keeping the total dose fixed). The colour bars show the approximate duration, $2\sigma$, of the dose. (c) Results for the high-affinity antibiotic, with parameters as in Table \ref{tab:1}. Colours are as in (b). (d) Time taken to recover from a step dose of antibiotic, as a function of the duration of the dose. ``Recovery'' is defined as an increase in $\lambda(t)$ to a value $0.9\lambda_0$, having previously been below this threshold. The recovery time is defined as the total time during which the growth rate is suppressed below $0.9\times\lambda_0$. The solid line shows results for the low-affinity antibiotic, the symbols show results for the high-affinity antibiotic. For the high-affinity case, the growth rate eventually recovers to its pre-dose value, on a timescale that is longer than shown in the plot.}
\label{fig:transientstepgauss}
\end{figure}

Real pharmacokinetic curves are of course smooth rather than step-like. To test the generality of our results, we also simulated the dynamical response to a smooth pulse of antibiotic, represented by a Gaussian function $a_{\rm ex}(t) = A\exp{[-(t-t_{\rm max})^2/(2\sigma^2)]}$, as illustrated in Figure \ref{fig:transientstepgauss}(a). We varied the peak dose intensity $A$ and the dose duration parameter $\sigma$, keeping the total dose, which is given approximately by $\sqrt{2\pi} A \sigma$, fixed at 4$\times$IC$_{50}$. For all plots shown in Fig. \ref{fig:transientstepgauss} we fix $t_{\rm max}=6$h.\\

Figure \ref{fig:transientstepgauss}(b) shows results for the low-affinity antibiotic. As for the step pulse, the bacterial growth rate is suppressed during the Gaussian antibiotic pulse, to an extent that depends on pulse intensity (the longer, less intense pulse shown by the red curve produces longer duration but weaker growth suppression than the shorter, more intense pulse shown by the blue curve). The model also predicts the same growth-rate overshooting phenomenon following the Gaussian antibiotic pulse which we observed for the step pulse. As for the step pulse, the magnitude of this overshoot increases as the drug-free growth rate $\lambda_0$ decreases (data not shown).\\

The response to a Gaussian pulse of a high-affinity antibiotic (Figure  \ref{fig:transientstepgauss}(c)) is also qualitatively similar to that for the step pulse (Figure \ref{fig:transientstep}(c)). As for the step pulse, for pulses of intensity below a threshold value, the growth rate recovers quickly following the antibiotic dose. However for pulses with intensity above the threshold, there is a post-antibiotic effect, in which growth suppression persists for long times after the antibiotic has been removed (recovery happens over timescales that are longer than those shown in Figure \ref{fig:transientstep}(c)). \\

Figure  \ref{fig:transientstepgauss}(d) shows the predicted recovery time after a Gaussian pulse of antibiotic, defined as the time to reach $\lambda=0.9\lambda_0$, as in Figure \ref{fig:transientstep}(d) for the step-like pulse. For the low-affinity antibiotic (solid line in Figure  \ref{fig:transientstepgauss}(d)), the time to recovery increases with the dose duration. This supports our prediction that for the low-affinity antibiotic, dose duration is the key determinant of treatment efficacy. For the high-affinity antibiotic (symbols in  Figure  \ref{fig:transientstepgauss}(d)), the time to recovery shows qualitatively similar behaviour to that for the step-like pulse (compare to Figure \ref{fig:transientstep}(d)), in that the time to recovery is very long for short, intense pulses, but decreases dramatically for pulses with intensity below a threshold. The more complex shape of the plot for low-intensity pulses, compared to the results for the step pulse (Figure \ref{fig:transientstep}(d)) is due to the shape of the Gaussian pulse; for values of $\sigma$ above $\sim 4.3$h, the growth rate no longer decreases below $0.9\lambda_0$. \\

 Taken together, our results for the Gaussian and step-like pulses of antibiotic show that the phenomena predicted by our model: (i) duration-dependent efficacy for ribosome-targeting antibiotics which bind with low affinity and/or are transported reversibly, (ii) growth-rate overshoot for these ``low-affinity'' antibiotics, (iii) peak intensity-dependent efficacy for ribosome-targeting antibiotics that bind with high affinity and/or are transported irreversibly and (iv) post-antibiotic effect for these ``high-affinity'' antibiotics, are all independent of the details of the form of the antibiotic dosage protocol.

\section*{Discussion}

In this paper, we have studied the dynamical response of the bacterial growth rate to sustained and transient antibiotic treatment, for ribosome-targeting antibiotics. The model that we have used is simple: it includes only antibiotic-ribosome binding, antibiotic transport, growth, and ribosome synthesis, with the latter two processes being dependent on the state of the system. In previous work \cite{greulich_2015}, this model has been shown to predict qualitatively different steady-state behaviour for two classes of ribosome-targeting antibiotics: ``low-affinity'' antibiotics which bind to ribosomes with low affinity and/or are transported reversibly across the cell boundary, and ``high-affinity'' antibiotics which bind with high affinity and/or are transported irreversibly.\\ 

Here, we have shown that these two classes of antibiotics also show qualitatively different dynamical responses to antibiotic treatment. Low-affinity antibiotics show a non-monotonic response, with a rapid decrease in growth rate upon exposure to antibiotic, followed by a slower partial recovery mediated by up-regulation of ribosome synthesis. Up-regulation of ribosome synthesis during exposure  also means that these antibiotics are predicted to show a striking growth rate overshoot upon removal of the antibiotic. In contrast, high-affinity antibiotics show a strongly concentration-dependent response: upon antibiotic exposure, the growth rate decreases very little if the antibiotic concentration is below a threshold given by the bifurcation point of the model dynamics, but it decreases almost to zero upon exposure to antibiotic concentrations above the threshold. Close to the threshold concentration the time taken to reach this maximal inhibition can, however, be very long: this behaviour can be understood by the fact that the dynamical trajectories of the model slow down as they pass close to the  location where the two fixed points have merged. Furthermore, the model predicts a pronounced post-antibiotic suppression of growth upon removal of a high-affinity antibiotic, for concentrations above the threshold -- a phenomenon that results from hysteresis in the model dynamics. \\

Are these predictions realistic? Of course many factors have not been included in the model. For example, we have assumed throughout that growth rate is determined by the active ribosome abundance, via Eq. (\ref{eq:1stGL}). Although this relation is well-established for steady-state growth, other factors may come into play during  transient growth-rate change. In particular, the growth rate may become limited by the supply of amino acids rather than by the abundance of free ribosomes. This  might reduce or eliminate the growth-rate overshoot predicted by our model for the low-affinity antibiotics. During an antibiotic pulse, when translation is inhibited, the total ribosome abundance is close to maximal ($r_{\rm tot} \approx r_{\rm max}$ in Eq. (\ref{eq:2ndGL})). According to the proteome partitioning model, this increased production of ribosomes comes at the expense of producing metabolic enzymes necessary for amino acid supply \cite{scott_science,scott_2014}. Thus, when the antibiotic is removed and ribosomes are released, there is likely to be a transient period when the rate of growth is limited by amino acid supply, before metabolic enzymes are re-synthesized to restore the balance between amino acid influx and the demands of translating ribosomes \cite{scott_2014}.  Our model also neglects any other effects of the antibiotics on bacterial physiology: for example, aminoglycosides may increase membrane permeability through the production of misfolded protein \cite{review_streptom_irr}. In addition, we do not model bacterial killing, either directly by antibiotic action, or indirectly via the body's immune system \cite{pankey}. Inclusion of these killing effects in the model would be likely to prevent the long-time recovery dynamics predicted here for the high-affinity antibiotics. \\

To conclusively assess the realism of the predictions reported here, one would need experimental tests. Several recently-developed bacterial growth techniques make such tests feasible. At the level of bulk cultures, continuous culture devices have been developed that allow measurement of growth rate during time-dependent antibiotic exposure  \cite{toprak_morbidostat}. At the level of individual cells, microfluidic devices in which the antibiotic concentration can be changed rapidly as growth is monitored in a microscope are also now possible  \cite{lin_2016}. The latter would be an especially interesting approach since the bistability which is manifested in our model for high-affinity antibiotics might lead to heterogeneous responses to antibiotic exposure among cells in a population. \\

If confirmed experimentally, the phenomena reported here would be of considerable clinical significance. In particular, our results make a clear prediction for the optimal pharmacodynamic strategy: for low-affinity drugs one should aim to maximise the time of exposure, while for high-affinity drugs, one should aim to maximise the peak dosage. Moreover, the latter are predicted to show a pronounced post-antibiotic effect, meaning that they can be effective for much longer than the actual duration of exposure. Post-antibiotic effects are a widely recognised, but poorly understood, pharmacodynamic phenomenon, and occur for various antibiotics including aminoglycosides \cite{mackenzie_1993,bundtzen_1981,isaksson_1988}. Our work suggests that models that  integrate molecular mechanism with bacterial cell physiology  can be a useful tool for understanding such clinically relevant growth inhibition phenomena and thus, potentially, for helping to improve clinical practice.\\

{\it Acknowledgments.}  PG and JD contributed equally to this work. This work was supported by the European Research Council under Consolidator Grant 682237-EVOSTRUC and by EPSRC under grant EP/J007404/1. PG was funded by a DFG research fellowship and RJA was funded by a Royal Society University Research Fellowship. JD was funded by a University of Edinburgh Physics and Astronomy Career Development Internship. MS was funded through an NSERC Discovery grant.\\

\section*{Appendix A: Bifurcation points of the model}
\begin{figure}
	\centering
	\includegraphics[width=0.9\linewidth]{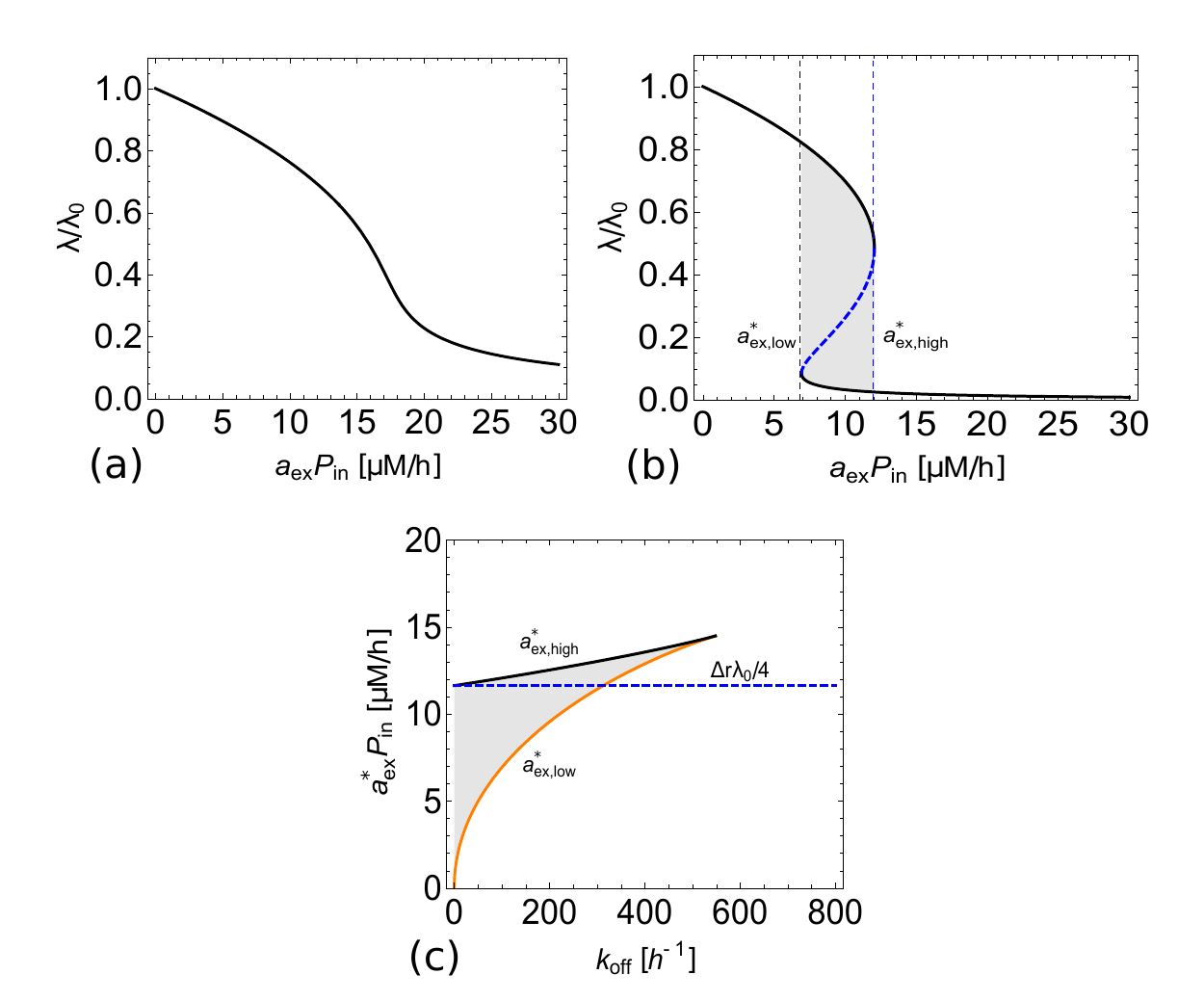}
	\caption{Illustration of the bifurcation points of the model. (a) Growth rate as a function of  $a_{\rm ex}P_{\rm in}$ for $k_{\rm off} = 1000 \, $h$^{-1}$, $P_{\rm in}=1$h$^{-1}$, $P_{\rm out} = 1\, $h$^{-1}$ and $k_{\rm on}=1000\mu$M$^{-1}$h$^{-1}$. For this parameter set the system has one fixed point.  (b) Growth rate as a function of $a_{\rm ex}P_{\rm in}$ for $k_{\rm off} = 100 \, $h$^{-1}$, $P_{\rm in}=1$h$^{-1}$, $P_{\rm out} = 1\, $h$^{-1}$ and $k_{\rm on}=1000\mu$M$^{-1}$h$^{-1}$. Here,  the system has regimes with one fixed point and a bistable regime with three fixed points (shaded area) -- two of which are stable (solid lines) and one unstable (dashed blue line). The bistable regime is bounded by the critical values $a^*_{\rm ex,low}$ and $a^*_{\rm ex,high}$. (c) The critical points $a^*_{\rm ex,low}$ and $a^*_{\rm ex,high}$, computed from the discriminant of Eq. (\ref{eq:Sn2}) as detailed in the text and plotted as function of $k_{\rm off}$. The bistable regime is shaded. The blue dashed line marks the analytical estimate $\Delta r \lambda_0/4$ for the upper critical point $a^*_{\rm ex, high}$, which is valid for $k_{\rm on} \gg \kappa_t$ and $P_{\rm out}k_{\rm off} \lesssim a_{\rm ex}P_{\rm in}\lambda/\Delta r$. Note that  since Eq. (\ref{eq:Sn2}) is symmetric under exchange of $k_{\rm off}$ and $P_{\rm out}$, $a^*_{\rm ex}P_{\rm in}$ as a function of $P_{\rm out}$ would look exactly the same for $k_{\rm off} =1\, $h$^{-1}$.}
	\label{fig:bifurcation}
\end{figure}

As we show in Fig.  \ref{fig:fixedpoints}, the model described by Eqs. (\ref{eq:antibDyn}) -- (\ref{eq:rbDyn}) may have different numbers of stationary points (stable and unstable fixed points), depending on the parameter values. Changes in the number and character of the fixed points of the model occur at critical   parameter values, and are known as bifurcation points. Fig.  \ref{fig:bifurcation} illustrates these bifurcation points in more detail. Here we use a parameter set intermediate between the low and high-affinity cases studied in the rest of the paper: $P_{\rm in}=1$h$^{-1}$, $P_{\rm out} = 1\, $h$^{-1}$, $k_{\rm on}=1000\mu$M$^{-1}$h$^{-1}$. We vary the parameter $k_{\rm off}$ and plot the  fixed points of the model as a function of $a_{\rm ex} P_{\rm in}$.  Fig.  \ref{fig:bifurcation}(a) shows results with $k_{\rm off} = 1000 \, $h$^{-1}$, for which there is  only a single fixed point. In contrast, for a smaller value of $k_{\rm off} = 100 \, $h$^{-1}$, as shown in Fig. \ref{fig:bifurcation}(b), for some values of $a_{\rm ex} P_{\rm in}$ the model has one fixed point (which is stable), and for other values of  $a_{\rm ex} P_{\rm in}$ there are three fixed points, two of which are stable and one unstable. The regime in which there are three fixed points is bistable (shaded) and is bounded by two bifurcation points, labelled $a^*_{\rm ex,low}$ and $a^*_{\rm ex,high}$ in  Fig. \ref{fig:bifurcation}(b), where the number of fixed points changes. The upper bifurcation point $a^*_{\rm ex,high}$ is associated with a steep decrease in the growth rate $\lambda$, since at this bifurcation point the upper stable fixed point is lost and $\lambda$ drops to the lower fixed point. This critical value  $a^*_{\rm ex,high}$ is very close to (though not exactly equal to) the IC$_{50}$.\\

The bifurcation points can be calculated by noting that the fixed points of the model dynamics are given by the roots of Eq. (\ref{eq:Sn2}). The number of roots --- and thus the number of fixed points -- is determined by the discriminant of Eq. (\ref{eq:Sn2}): if the discriminant is positive there are three roots, otherwise, there is only one root. Thus the zeros of the discriminant mark the bifurcation points. Since Eq. (\ref{eq:Sn2}) is a cubic equation in $\lambda$, it may be written as $a\lambda^3 + b\lambda^2 +c\lambda + d$, with discriminant $\Delta = b^2 - 4ac^3 - 4b^3d -  27a^2d^2 + 18abcd$. The zeros of the discriminant can be computed numerically. Fig. \ref{fig:bifurcation}(c) shows the results of such a computation: here the bifurcation points $a^*_{\rm ex,low}$ and $a^*_{\rm ex,high}$ are plotted as a function of $k_{\rm off}$. Since the discriminant itself is cubic in $a_{\rm ex}$ it may have either one or three zeros; those at positive  $a_{\rm ex}$ correspond to  $a^*_{\rm ex,low}$ and $a^*_{\rm ex,high}$  \footnote{There is also one unphysical zero point of the discriminant at negative values of $a_{\rm ex}$ not shown here.}. For low values of $k_{\rm off}$ there are two  bifurcation points as in Fig. \ref{fig:bifurcation}(b), while for high values of $k_{\rm off}$ there is no bifurcation, as in Fig. \ref{fig:bifurcation}(a).\\

We can also obtain an analytical estimate for the upper bifurcation point $a_{\rm ex}=a^*_{\rm ex,high}$, which corresponds to the antibiotic concentration at which the model predicts a threshold drop in growth rate. To this end, we rewrite Eq. (\ref{eq:Sn2}) in the form
\begin{align}\label{eq9}
0 &=\frac{\lambda}{k_{\rm off}P_{\rm out}}\left(\frac{a_{\rm ex}P_{\rm in}}{\Delta r} + \lambda - \frac{1}{\lambda_0} \lambda^2\right) \\ \nonumber
&+ \left(\frac{\kappa_t}{k_{\rm on}}\right)\left[1 + \left(\frac{1}{P_{\rm out}} + \frac{1}{k_{\rm off}} - \frac{1}{\lambda_0}\right) \lambda -  \left(\frac{1}{P_{\rm out}\lambda_0} + \frac{1}{k_{\rm off}\lambda_0}  - \frac{1}{P_{\rm out}k_{\rm off}} \right) \lambda^2 - \frac{1}{P_{\rm out}k_{\rm off}\lambda_0} \lambda^3 \right],
\end{align}
(note that we have multiplied Eq. \ref{eq:Sn2} by a factor of 4). Interestingly, this equation depends only on the combination $a_{\rm ex} P_{\rm in}$ rather than on $a_{\rm ex}$ and $P_{\rm in}$ independently. This is why we have used the parameter combination $a_{\rm ex} P_{\rm in}$ in  Fig. \ref{fig:bifurcation}; it also implies that the critical values  $a^*_{\rm ex}$ scale as $a^*_{\rm ex} \sim 1/P_{\rm in}$. For  $k_{\rm on} \gg \kappa_t$, and $k_{\rm off}P_{\rm out}$ not too large, the second term in Eq. (\ref{eq9}) can be neglected and we arrive at the quadratic equation
\begin{align}
\label{eq:cubic_alt}
0 = \frac{a_{\rm ex} P_{\rm in}}{\Delta r} + \lambda + \frac{1}{\lambda_0} \lambda^2. 
\end{align}
The zero of the discriminant is then at
\begin{align}
a^*_{\rm ex} P_{\rm in} = \frac{\Delta r \lambda_0}{4}.
\end{align}
Setting $r_{\rm min} = 19.3 \mu$M, $r_{\rm max} = 65.8 \mu$M, and $\lambda_0 = 1\,$h$^{-1}$, as in Table \ref{tab:1}, this gives $a^*_{\rm ex} P_{\rm in} =  11.625 \mu$M\,h$^{-1}$, shown in Fig. \ref{fig:bifurcation}(c) as the blue dashed line. Fig. \ref{fig:bifurcation}(c) shows that for small values of $k_{\rm off}$ this provides a very good estimate for the upper bifurcation point $a^*_{\rm ex,high}$. Thus, $a^*_{\rm ex} = 11.625\mu$M\,h$^{-1}/P_{\rm in}$ is a good estimate for the threshold antibiotic concentration and the IC$_{50}$ for high-affinity antibiotics. Remarkably, this approximation does not explicitly depend on $P_{\rm out}$, $k_{\rm off}$ or $k_{\rm on}$. For large values of $k_{\rm off}$ or $P_{\rm out}$, however, this approximation does not hold anymore, since the prefactor $1/(k_{\rm off}P_{\rm out})$ in Eq. (\ref{eq9}) decreases the importance of the first term relative to the second term.

\section*{Appendix B: Analytical calculation of inhibition time for a high-affinity antibiotic using the adiabatic approximation}

Incorporating expression (\ref{eq:SynthRate}) for the ribosome synthesis rate into the  dynamical equations (\ref{eq:antibDyn})-(\ref{eq:rbDyn}), our model can be expressed as:
\begin{equation}\label{app:eq1}
	\dot{a}=-\lambda a-k_{\rm on}(r_{\rm u}-r_{\rm min})a+P_{\rm in}a_{\rm ex}-P_{\rm out}a+k_{\rm off}r_{\rm b}
\end{equation}
\begin{equation}\label{app:eq2}
\dot{r}_{\rm u}=-\lambda r_{\rm u}-k_{\rm on}(r_{\rm u}-r_{\rm min})a+\lambda(r_{\rm max}-\lambda c)+k_{\rm off}r_{\rm b}
\end{equation}
\begin{equation}\label{app:eq3}
\dot{r}_{\rm b}=-\lambda r_{\rm b}+k_{\rm on}(r_{\rm u}-r_{\rm min})a-k_{\rm off}r_{\rm b}
\end{equation}
where $\Delta =r_{\rm max}-r_{\rm min}$, $c\equiv\Delta r(\frac{1}{\lambda_0}-\frac{1}{\kappa_t \Delta r})$ and $\lambda$ is a function of $r_{\rm u}$ via Eq. (\ref{eq:1stGL}). Making the adiabatic approximation, {\em i.e.} setting  $\dot{a}=0$, and using Eq. (\ref{eq:1stGL}) to eliminate $r_{\rm u}$, Eq. (\ref{app:eq1}) gives 
\begin{equation}\label{app:eq4}
	a=\frac{P_{\rm in}a_{\rm ex}+k_{\rm off}r_{\rm b}}{P_{\rm out}+\lambda(1+\frac{k_{\rm on}}{\kappa_t})}.
\end{equation}

Substituting Eq. (\ref{app:eq4}) into Eqs. (\ref{app:eq2}) and  (\ref{app:eq3}) and using Eq. (\ref{eq:1stGL}) to change variables from $r_{\rm u}$ to $\lambda$, we obtain
\begin{equation}\label{app:eq5}
	\dot{\lambda}=-\lambda^2(1+c\kappa_t)-\frac{k_{\rm on}\lambda(P_{\rm in}a_{\rm ex}+k_{\rm off}r_{\rm b})}{P_{\rm out}+\lambda(1+\frac{k_{\rm on}}{\kappa_t})}+k_{\rm off}\kappa_t r_{\rm b}+\Delta r \lambda \kappa_t
\end{equation}
and 
\begin{equation}
	\dot{r}_{\rm b}=-\lambda r_{\rm b}+k_{\rm on}\lambda/\kappa_t\frac{P_{\rm in}a_{\rm ex}+k_{\rm off}r_{\rm b}}{P_{\rm out}+\lambda(1+\frac{k_{\rm on}}{\kappa_t})}-k_{\rm off} r_{\rm b} .
\end{equation}
In Eq. (\ref{app:eq5}), we then make the approximation that $\lambda(1+k_{\rm on}/\kappa_t)\gg P_{\rm out}$ \footnote{We expect this approximation to be valid close to the upper fixed points, whose bifurcation we are concerned with here. Close to the lower stable fixed point, where $\lambda \to 0$, the approximation may not hold. }. Since we are dealing with a high-affinity antibiotic, we also set $k_{\rm off} = 0$. This allows us to express the model as an equation in one variable only (the growth rate $\lambda(t)$):
\begin{equation}\label{app:eq6}
	\dot{\lambda}=-\lambda^2(1+c\kappa_t)-\frac{k_{\rm on}P_{\rm in}a_{\rm ex}}{(1+\frac{k_{\rm on}}{\kappa_t})}+\Delta r \lambda \kappa_t .
\end{equation}

Returning to Eq. (\ref{app:eq6}), we can integrate the trajectory $\lambda(t)$ to predict the time $T_c$ required for $\lambda$ to reach a predefined threshold $\lambda_c$:
\begin{equation}
T_c=\int_{\lambda=\lambda_0}^{\lambda=\lambda_c}\frac{d \lambda}{-\lambda^2(1+c\kappa_t)-\frac{k_{\rm on}P_{\rm in}a_{\rm ex}}{(1+\frac{k_{\rm on}}{\kappa_t})}+\Delta r \lambda \kappa_t}.
\end{equation}
This integral can be solved using the substitution $u = \frac{\Delta r \kappa_t/2-(1+c \kappa_t) \lambda}{\sqrt{(1+c \kappa_t) C - (\Delta r \kappa_t)^2/4}}$, where we have defined $C=\frac{k_{\rm on}P_{\rm in}a_{\rm ex}}{1+\frac{k_{\rm on}}{\kappa_t}}$. This gives the following result:
\begin{eqnarray}
\label{app:eq7}
	T_c&=&\frac{1}{\sqrt{(1+c \kappa_t)C-(\Delta r \kappa_t)^2/4}}\int_{u(\lambda_0)}^{u(\lambda_c)}du\frac{1}{1+u^2}
\\ \nonumber
&=&\frac{1}{\sqrt{(1+c \kappa_t)C-(\Delta r \kappa_t)^2/4}}\Bigg(\arctan\bigg(\frac{\kappa_t \Delta r/2-(1+c\kappa_t)\lambda_c}{\sqrt{(1+c \kappa_t)C-(\Delta r \kappa_t)^2/4}}\bigg) -\arctan\bigg(\frac{\kappa_t \Delta r/2-(1+c\kappa_t)\lambda_0}{\sqrt{(1+c \kappa_t)C-(\Delta r \kappa_t)^2/4}}\bigg)\Bigg)
\end{eqnarray}
Setting $\lambda_c=0.01\lambda_0$ in Eq. (\ref{app:eq7}) leads to the results shown as the solid curve in Figure \ref{fig:fullinhibition}.  Note that this integration is  only valid if $(\Delta r \kappa_t)^2 - 4 (1 + c \kappa_t) C < 0$. Otherwise, the denominator becomes zero and $T_c$ diverges. This would be the case if the system is at the upper stable fixed point of the dynamics, such that the growth rate is not significantly decreased upon exposure to the antibiotic.\\

\begin{figure}
	\centering
	\includegraphics[width=0.9\linewidth]{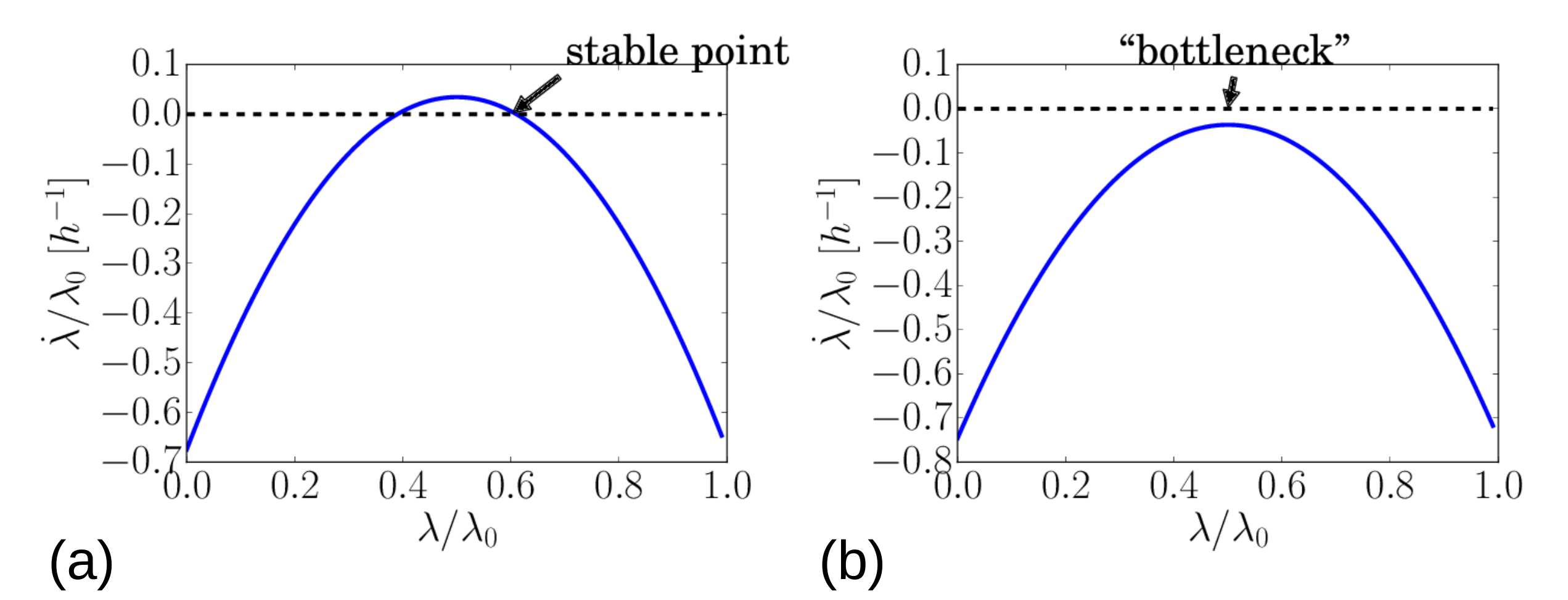}
	\caption{Rate of change of growth rate $\dot{\lambda}$, plotted as a function of $\lambda$, as given by the adiabatic approximation, Eq. (\ref{app:eq6}). Panel (a) shows results for $a_{\rm ex}=0.95\times$IC$_{50}$ (just below the bifurcation point of Figure \ref{fig:fixedpoints}(b)), while panel (b) shows results for $a_{\rm ex}=1.05\times$IC$_{50}$ (just above the bifurcation point). }
	\label{fig:philip}
\end{figure}

This analysis also allows us to understand the origin of the very slow inhibition dynamics for the high-affinity antibiotic, for values of $a_{\rm ex}$ just above the bifurcation point, as shown in Figure \ref{fig:trajectories_irrev}(a). Figure \ref{fig:philip} shows the rate of change of the growth rate, $\dot{\lambda}$, plotted as a function of $\lambda$, as predicted by Eq.  (\ref{app:eq6}), for the high-affinity parameter set. Figure \ref{fig:philip}(a) shows results for $a_{\rm ex}=0.95\times$IC$_{50}$ (just below the bifurcation point): the fixed points correspond to zeroes of  $\dot{\lambda}$ and the stable one is indicated by the arrow (there is of course also another stable fixed point at very small $\lambda$, but this is lost in the quadratic approximation of Eq. (\ref{app:eq6})). Figure \ref{fig:philip}(b) shows equivalent results for a slightly higher antibiotic concentration, $a_{\rm ex}=1.05\times$IC$_{50}$, just above the bifurcation point. Here the fixed points are lost, but the rate of change of $\lambda$ still comes close to zero, implying that the speed of inhibition by the antibiotic will be very slow. This slow dynamics close to the bifurcation point can be thought of as a ``bottleneck'' in the inhibition trajectory. 

\end{document}